\documentclass[published]{JHEP3} 
\JHEP{09(2003)070}
\received{May 9, 2003}
\accepted{September 29, 2003}

\skip\footins = 1\bigskipamount plus 2pt minus 4pt

\usepackage{epsfig}

\newcommand{\GeV}{{\rm \,GeV}}
\newcommand{\GeVsq}{{\rm \,GeV}^2}
\newcommand{\pb}{\mbox{${\rm \,pb}$}}
\newcommand{\als}{\alpha_s}

\newcommand{\gsim}{\gtrsim}

\newcommand{\sigmab}{\ensuremath{\sigma_{\rm b}}}
\newcommand{\sigmabj}{\ensuremath{\sigma_{\rm b,jet}}}
\newcommand{\sigmabjm}{\ensuremath{\sigma_{\rm b,jet,muon}}}
\newcommand{\sigmabm}{\ensuremath{\sigma_{\rm b, muon}}}

\newcommand{\bmeson}{\mbox{$B$-{meson}~}}
\newcommand{\bmesonx}{\mbox{$B$-{meson}}}
\newcommand{\bquark}{\mbox{$b$-{quark}~}}
\newcommand{\bquarks}{\mbox{$b$-{quarks}~}}
\newcommand{\bquarkx}{\mbox{$b$-{quark}}}
\newcommand{\etb}{\mbox{$E_{T,{\rm b}, {\rm Breit}}$}}
\newcommand{\etab}{\mbox{$\eta_{{\rm b}, {\rm lab}}$}}
\newcommand{\etjet}{\mbox{$E_{T,{\rm jet}, {\rm Breit}}$}}
\newcommand{\etajet}{\mbox{$\eta_{{\rm jet}, {\rm lab}}$}}
\newcommand{\etamujet}{\mbox{$\eta_{{\rm muonjet}, {\rm lab}}$}}
\newcommand{\etmu}{\mbox{$E_{T,\mu, {\rm lab}}$}}
\newcommand{\pmu}{\mbox{$P_{\mu, {\rm lab}}$}}
\newcommand{\etamu}{\mbox{$\eta_{\mu, {\rm lab}}$}}

\makeatletter

\newbox\JHEP@outputbox
\gdef \@makecol {%
   \ifvoid\footins
     \setbox\@outputbox \box\@cclv
   \else
     \setbox\@outputbox \vbox {%
       \boxmaxdepth \@maxdepth
       \@tempdima\dp\@cclv
	\unvbox \@outputbox
       \unvbox \@cclv
       \vskip-\@tempdima
     }%
     \setbox\JHEP@outputbox \vbox {%
       \vskip \skip\footins
       \color@begingroup
         \normalcolor
         \footnoterule
         \unvbox \footins
       \color@endgroup
       }%
   \fi
   \xdef\@freelist{\@freelist\@midlist}%
   \global \let \@midlist \@empty
   \@combinefloats
   \ifvbox\@kludgeins
     \@makespecialcolbox
   \else
     \setbox\@outputbox \vbox to\@colht {%
       \@texttop
       \dimen@ \dp\@outputbox
       \unvbox \@outputbox
       \vskip -\dimen@
       \unvbox\JHEP@outputbox
       \@textbottom
       }%
   \fi
   \global \maxdepth \@maxdepth
}
\makeatother

\title{Definition and calculation of bottom quark cross-sections in deep-inelastic scattering at HERA and determination of their uncertainties}

\author{Tancredi Carli,$^{a}$ Vincenzo Chiochia$^{bcd}$
 and Katarzyna Klimek$^{c}$ \\
\llap{$^a$}EP Division, CERN, 1211 Geneva, Switzerland\\
\llap{$^b$}DESY, Notkestr. 85, 22607 Hamburg, Germany\\
\llap{$^c$}University of Hamburg, Luruper Chaussee 149, 22607 Hamburg, Germany\\
\llap{$^d$}Physik Institut der Universit\"at Z\"urich-Irchel, 8057 Z\"urich, Switzerland\\
E-mail: \email{Tancredi.Carli@cern.ch},
 \email{Vincenzo.Chiochia@cern.ch},
 \email{kklimek@mail.desy.de}}
                                 
\abstract{The uncertainties involved in the calculation of bottom
  quark ($b$-quark) cross-sections in deep-inelastic scattering at
  HERA are studied in different phase space regions.  Besides the
  inclusive $b$-quark cross-section, definitions closer to the
  detector acceptance requiring at least one high energetic muon from
  the semi-leptonic $b$-quark decay or a jet with high transverse
  energy are investigated.  For each case the uncertainties due to the
  choice of the renormalisation and factorisation scales as well as the
  $b$-quark mass are estimated in the perturbative NLO QCD calculation
  and furthermore uncertainties in the fragmenation of the $b$-quark
  to a $B$-{meson} and in its semi-leptonic decay are discussed.}

\begin{document} 

\section{Introduction}

Quantum Chromodynamics (QCD) has proven to give a correct description
of the strong interaction, if a hard scale is involved in the process
such that perturbation theory is applicable.  The large charm ($c$)
and especially the bottom ($b$) quark mass ($m_b$) provide such a hard
scale, since $m_b \gg \Lambda$, where $\Lambda \approx 250\, {\rm MeV}$ is the typical QCD
scale.  The large mass screens the collinear singularities, so that
there is no need to apply a jet algorithm or to subtract them into a
fragmentation function in the perturbative cross section calculation.
Non-perturbative contributions to heavy quark momentum distributions
are of order ${\cal O}(\Lambda/m_b)$.  The production of bottom quarks
should therefore be well described by perturbative QCD calculations,
but the inclusion of non-perturbative effects is unavoidable to get a
physically complete result.

The \bquark production cross-sections in strong interactions have been
measured in proton-antiproton collisions at SPS~\cite{pl:b256:121} and
at Tevatron~\cite{prl:71:500}--\cite{pl:b487:264} and, more recently,
in two-photon interactions at LEP~\cite{pl:b503:10} and in
photon-proton collisions ($\gamma p$) at the electron-proton ($ep$)
collider HERA~\cite{pl:b467:156,epj:c18:625}.  For all these
measurements, except the early SPS data~\cite{pl:b256:121}, the
measured \bquark production cross-sections lie above perturbative QCD
expectations calculated up to next-to-leading order (NLO) in the
strong coupling, $\als$.  Since the production of bottom quarks is an
important background process in searches for phenomena beyond the
Standard Model at present and future high energy colliders, the
understanding of the observed excess of the measurement over the
perturbative QCD expectation, has recently been lively
discussed. Possible proposed solutions involve higher order effects
estimated by resummation of large
logarithms~\cite{Cacciari:1998it,Ball:2001pq}, new parton evolution
schemes~\cite{Jung:2001rp,Hagler:2000dd,Lipatov:2001ny}, fragmentation
effects~\cite{Cacciari:2002pa} or even the exchange of low-mass
supersymmetric particles~\cite{Berger:2000mp}.  Meanwhile the last
possibility seems to be excluded from LEP data~\cite{Janot:2003cr}.
 
Deep-inelastic scattering (DIS) offers the unique opportunity to study
the production mechanism of bottom quarks in a particularly clean
environment where a point-like projectile, a virtual photon
($\gamma^*$), collides with a proton.  It will be interesting to see,
if the excess observed in
photoproduction~\cite{pl:b467:156,epj:c18:625}, where the photon is
real and can develop a hadronic structure, persists for virtual
photons.  The large centre-of-mass energy of HERA colliding $27.5$\GeV\
electrons on $920$\GeV\ protons allows a sufficient number of
$b$-$\bar{b}$ pairs to be produced and provides therefore an excellent
and clean testing ground.

The calculation of the \bquark production cross-section requires the
knowledge of the parton densities (mainly the one of the gluon), of
the perturbatively calculable hard parton parton subprocess and the
correct modeling of the long range effects binding the \bquark in the
hadrons.  The Feynman diagram of the basic production meachanism for
the reaction $\gamma^* p \to b \bar{b} X$ is sketched in leading order
(LO) of the strong coupling ${\cal O}(\alpha_s)$ in
figure~\ref{fig:b_feyn}$a$.  As examples some generic Feynman diagrams
of the NLO contributions are shown in figure~\ref{fig:b_feyn}$b--d$.
The dominant process is the fusion of the virtual photon emitted by
the electron with a gluon from the proton.  The production of \bquarks
is therefore directly sensitive to the gluon density in the
proton. Furthermore the gluon evolution can be tested in the presence
of several possible hard scales like $m_b$, the photon virtuality
$Q^2$ or the transverse energy of the jet initiated by the bottom
quark.

A recent theoretical review on the NLO QCD calculation of heavy quark
cross-section in DIS can be found in~\cite{vanNeerven:2001tb}.

\FIGURE[t]{
\begin{picture}(0,0) 
 \put(0,-5){{($a$)}} 
 \put(100,-5){{($b$)}}
 \put(190,-5){{($c$)}}
 \put(280,-5){{($d$)}} 
\end{picture}
\epsfig{file=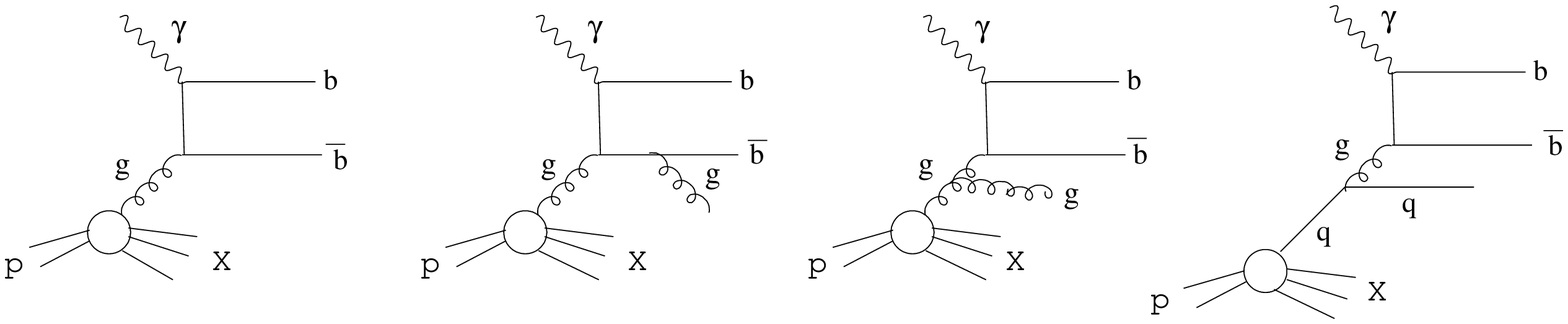,width=14.cm}%
\caption{($a$) Leading order Feynman diagram for the reaction $\gamma p
  \to b \bar{b} X$, ($b$ $c$ $d$) some generic Feynman diagrams of the
  ${\cal O}(\alpha_s^2)$ contributions.\label{fig:b_feyn}}}

In this paper we investigate the uncertainties involved in the
calculation of the bottom quark production cross-section in DIS for
different cross-section definitions and we address the question how
the calculations can be best compared to experimental results: a short
description of the perturbative NLO QCD calculation based on the
HVQDIS program~\cite{pr:d57:2806,np:b452:109,pl:b353:535}, the model
of the fragmentation of the \bquark in a \bmeson and its semi-leptonic
decay to a muon and a jet.  Tools to simulate \bquark production are
described in section~\ref{sec:mcgen}.  After a short overview of the
experimental techniques to identify bottom quarks, different
cross-section definitions based on the measurable quantities: the
scattered electron, the muon from the semi-leptonic \bquark decay and
the jet initiated by the \bquark are presented in
section~\ref{sec:cross_def}.  Section~\ref{sec:pqcd_cal} presents the
necessary elements needed for the calculation of the cross-section for
the reaction $e p \to b \bar{b} \to e \, \mu \, jet \, X$.  The LO QCD
calculation based on HVQDIS is compared to the LO QCD models Monte
Carlo simulations in section~\ref{sec:lo_calc}.  The uncertainties
involved in the NLO QCD cross-section calculations are discussed in
section~\ref{sec:uncertain}.  This includes the uncertainties of the
factorisation and the renormalisation scales, the \bquark mass, the
parton densities, the phenomenological description of the
fragmentation process and other non-perturbative effects.  In
addition, the uncertainties involved in extrapolating the measurable
cross-section $e p \to b \bar{b} \to e \, \mu \, jet \, X$ to a more
inclusive cross-section like e.g.\ $e p \to b \bar{b} \to e \, X$ are
discussed in section~\ref{sec:model}.
 
\section{Leading order QCD Monte Carlo simulation programs}
\label{sec:mcgen}

QCD Monte Carlo models are able to simulate the hadronic final state
of a DIS event in full detail. The four-momenta of all produced
particles are made explicitly available.  These programs are
indispensable tools to correct the experimental data for detector
effects.  Although they only contain the leading order matrix elements
of the hard subprocess, they include higher orders in a leading
logarithmic approximation and they make use of models providing
detailed treatment of the non-perturbative fragmentation phase. Here,
there are mainly used to test the simple fragmentation model
implemented in the NLO QCD calculation and to study the uncertainty of
extrapolating measured cross-section to more inclusive ones.  Another
purpose is to study possible effects from higher orders implemented in
the leading logarithm approximation.

The RAPGAP Monte Carlo~\cite{rapgap} incorporates the ${\cal O}
(\alpha_s)$ QCD matrix elements (ME) and approximates higher order
parton emission using the concept of parton showers
(PS) \cite{partonshower} based on the leading logarithm DGLAP
equations~\cite{dglap1,dglap2,dglap3}.  QCD radiation can occur before
and after the hard subprocess. The formation of hadrons is simulated
using the LUND string model~\cite{lund} as implemented in
JETSET~\cite{jetset}.  As an option parton showers can also be
simulated using ARIADNE~\cite{ariadne}. In this model gluon emissions
are treated by the colour dipole model~\cite{cdm1,cdm2} (CDM) assuming
a chain of independently radiating dipoles spanned by colour connected
partons.

In the CASCADE Monte Carlo simulation program~\cite{cascade} higher
order parton emissions based on the CCFM~\cite{ccfm1}--\cite{ccfm4}
evolution equations are matched to a ${\cal O}(\alpha_s)$ matrix
element~\cite{lome_hq1}--\cite{Schuler:1988wj}, where the incoming
parton can be off-shell.  The CCFM evolution equations are based on
$k_T$-factorisation and angular ordering which is a consequence of
colour coherence, i.e.\ due to the interference properties of the
radiated gluons.  As a result in the appropriate limit they reproduce
the DGLAP~\cite{dglap1,dglap2,dglap3} and the
BFKL~\cite{bfkl1,bfkl2,bfkl3} approximation.  At small values of the
parton momentum fractions $z$, a random walk of the transverse parton
momenta $k_T$ is obtained.  The CCFM evolution is based on
$k_T$-factorisation~\cite{ktfact}, where the partons entering the hard
scattering matrix element are free to be off-shell, in contrast to the
collinear approach (DGLAP) which treats all partons entering the hard
subprocess as massless.  Off-shell matrix elements of heavy flavour
lepto- and hadroproduction processes have been calculated in
ref.~\cite{ktfact}.  The unintegrated gluon
density\footnote{Unintegrated means that the gluon density depends on
  the transverse parton momenta emitted along the cascade. This
  dependence is not integrated out as in the DGLAP approach where the
  gluon density only depends on the energy fraction $x$ and on the
  squared transverse momentum transfer.}  used in CASCADE is extracted
through a fit to the proton structure function $F_2$ measured at
HERA~\cite{Aid:1996au}.  The fit was performed in the range
$x<10^{-2}$ and $Q^2 > 5\GeVsq$.  Recently, it has been shown that
CASCADE is able to correctly reproduce the \bquark production
cross-section in $p\bar{p}$-collisions at Tevatron~\cite{Jung:2001rp}.

In all calculations no QED corrections have been included.

\section{Measurement and definition of bottom quark cross-sections}
\label{sec:cross_def}

When a \bquark with high transverse energy is produced in a hard
strong interaction, a jet can be measured in the final state. The jet
is composed by a hadron containing the \bquark (in most cases a
\bmesonx) and other hadrons produced in the fragmentation process.
Due to the large \bquark mass the jet is expected to be wider than a
jet at similar transverse energy initiated by a light quark.  However,
since the rate of \bquarks is much smaller than the one of light
quarks, this characteristic feature is not sufficient to clearly
identify the presence of a \bquark.  Therefore usually the
semi-leptonic decay modes of the $B$-meson to electrons and to muons
$B \to e X$ and $B \to \mu X$ are exploited to identify the presence
of a \bquarkx. This largely suppresses the background from the light
quarks $u$, $d$ and $s$, which basically only can fake such a
signature when a hadron produced in a light quark jet is misidentified
as an electron or muon. The remaining background from charm quarks can
be statistically separated from the \bquark signal, since the lifetime
of the meson containing the charm quark and the kinematics of the
meson decay is significantly different.  This is due to the smaller
charm mass.  In many cases the analysis of the muon channel is
experimentally easier than the one of the electron channel. Therefore,
here we only refer to the muon channel.\footnote{ All the arguments
  are, of course, also valid for semi-leptonic decays into
  electrons.}

The experimental method often relies on the measurement of the
transverse momentum ($P_T^{rel}$) of the muon produced in the
semi-leptonic decay relative to the jet axis.  Due to the higher
\bquark mass the spectrum of this quantity is harder for jets
initiated by \bquarks than by charm quarks. When this method is used,
a jet and a muon have to be measured.  Complementary information can
be obtained from lifetime measurements of the \bmeson with the help of
high precision silicon vertex detectors.  A particularly elegant and
simple method measures only the impact parameter\footnote{The impact
  parameter is the distance of closest approach of the muon track to
  the primary event vertex.  The average impact parameter measured in
  the laboratory frame is directly related to the lifetime of the
  \bmeson in its rest frame.}  of the muon track with respect to the
main interaction vertex.  Therefore, in principle in this method only
a muon is needed to extract the \bquark production cross-section.  An
alternative \bquark tag can be obtained via the identification of a
$D^*$-meson.  This method gives access to low \bquark momenta, but is
not considered here, since the efficiency of this method is rather
low.

Besides the presence of the scattered electron the experimental
definition of a \bquark \linebreak cross-section in DIS therefore
requires the presence of a muon with typically a momentum bigger than
a few GeV and/or the presence of a jet associated to the muon at high
transverse energy in the detector.  The \bquark cross-section
measurements should always be quoted for the reaction $ep \to e \mu
(jet) X$.

To circumvent a complex cross-section definition, to facilitate the
evaluation of the NLO QCD calculation and to make a comparison between
different experimental results easier, in the past often extrapolations of the
experimentally measured cross-section to a more inclusive definition
have been made using leading order QCD Monte Carlo simulation
programs. Such a procedure can only be adopted, if the models used for
the extrapolation is reliable. Since in many cases the involved
extrapolations are large, it is doubtful whether the accuracy of the
leading order QCD Monte Carlo simulation programs is sufficient.  In
any case, unnecessary model uncertainties are introduced in the
experimental cross-section measurement which, when not properly
stated, can make the direct comparison of data very difficult, if the
extrapolation have been based on different model
assumptions.\footnote{This is almost always the case, if older data
  are compared to recent ones.}

Besides the experimentally required cross-section definition for the
reaction $ e p \to b \bar{b} \to e \, \mu \, jet X$ ($\sigma_{\rm
  b,jet,muon}$), in this paper we will investigate the total inclusive
\bquark cross-section only defined by the scattered electron measured
in the detector ($\sigma_{\rm b}$), the \bquark \linebreak
cross-section where in addition a hard jet is measured ($\sigma_{\rm
  b,jet}$) and the \bquark cross-section where a muon, but no hard jet
is required ($\sigma_{\rm b,muon}$).
 
The scattered electron should be well measured in the main part of the
detector.  Therefore we require that the minimal $Q^2$ of the
exchanged photon should be $Q^2 > 2 \GeVsq$.  Since with the present
HERA data sample very high $Q^2$ values are not accessible, $Q^2$ is
restricted to $Q^2 < 1000 \GeVsq$.  The quantity $y = Q^2/(s x)$,
where $x = Q^2/(2 P q)$ is the Bjorken scaling variable, $q$ (P) is
the four-momentum of the exchanged photon (the proton beam) and $s$ is
the squared centre-of-mass energy, is set to $0.05 < y <0.7$. The
lower cut on $y$ ensures that the squared invariant mass of the
hadronic final state $W^2 = (q + P)^2 = y s$ is large enough to
produce a final state with high transverse energy.  The higher cut on
$y$ is needed to ensure a good reconstruction of the event kinematics
and to suppress non-DIS background.\footnote{High values of $y$
  correspond to low energies of the scattered electron where the
  unambiguous identification of the scattered electron becomes more
  difficult.}

The jet should be well inside the acceptance of the calorimeter and
the muon well inside the acceptance of the muon chambers and the inner
tracking detectors.  We require here $-2 < \etajet < 2.5$ and
$30^{\circ} < \theta_{\mu,{\rm lab}} < 160^{\circ}$ which corresponds to $-1.7 <
\eta_{\rm \mu, lab} < 1.3$. When the muon should be identified in the
muon chambers it needs a sufficiently large momentum to penetrate the
calorimeters.  In our case we consider $\pmu > 2 \GeV$.

A hard interaction is signaled by the presence of a jet with high
transverse energy.  The jet definition is based on the inclusive
$k_T$-algorithm~\cite{invkt,ktclus}.  To cluster particles or partons
to jets the $E_T$ scheme is used. The resulting jet four-momentum is
therefore massless.  This algorithm is well suited to make
quantitative comparisons with NLO calculation, since it is infra-red
and collinear safe~\cite{Seymour:1998kj}.  To remove the purely
kinematic dependence of the transverse jet energy on the photon
virtuality $Q^2$, the transverse jet energy is measured in a frame
where the photon and the proton collide head on.  Such a frame is
e.g.\ the Breit frame defined by $2 x \vec{P}+ \vec{q} = 0$.  A jet
with a transverse energy $\etjet > 6 \GeV$ is required in the Breit
frame.  The jet should be within the detector acceptance. We require
here: $-2 < \etajet < 2.5$.

The different cross-section definitions are summarised in
table~\ref{tab:cross-section_def}.

\TABLE[t]{ 
\begin{tabular}{|l|ccc|}
\hline
       & DIS & JET & MUON \\ 
       & $2 < Q^2 < 1000 \GeVsq$ & $\etjet > 6 \GeV$ & $\pmu > 2 \GeV$ \\
Symbol & $0.05 < y < 0.7 $ & $-2 < \etajet < 2.5$ & $30^o < \theta_\mu < 160^o$ \\
\hline 
\sigmab   & x &   &   \\ 
\sigmabj  & x & x &   \\ 
\sigmabm  & x &   & x \\ 
\sigmabjm & x & x & x \\\hline
\end{tabular}
\caption{Definition of the \bquark production cross-sections used in
  this analysis.  For each definition the applied cut is marked with
  the symbol $x$.\label{tab:cross-section_def}}}

\section{Perturbative LO and NLO QCD calculations and fragmentation models}
\label{sec:pqcd_cal}

\FIGURE[t]{\epsfig{file=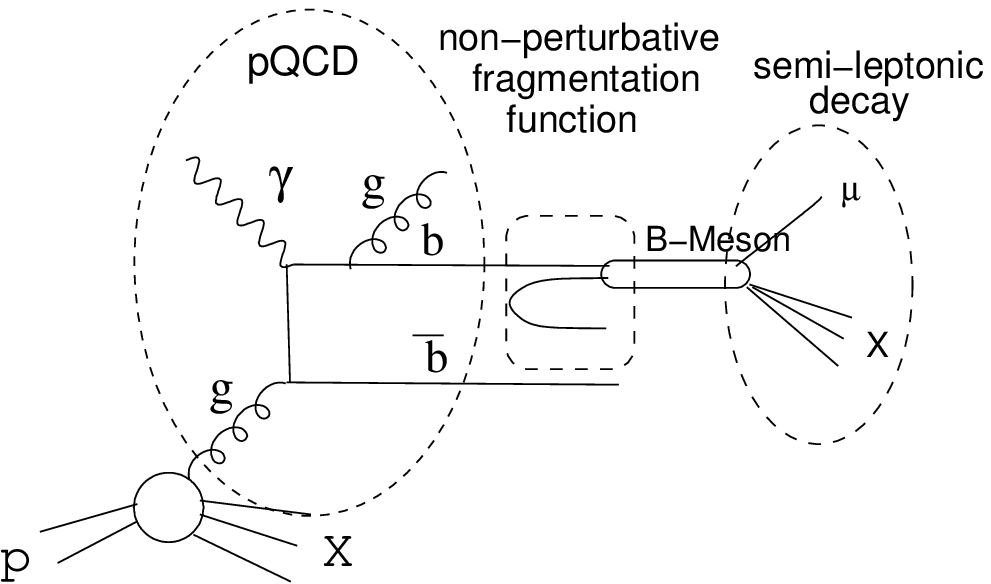, width=.6\textwidth}%
\caption{Sketch of the production of a \bmeson in DIS and its
  subsequent decay.\label{fig:b_frag}}}

The \bquark cross-section measurements often rely on the
identification of a muon from the semi-leptonic decay of a
\bmesonx. To be comparable to the measurements the calculation of the
\bquark cross-section proceeds in three steps as it is illustrated in
figure~\ref{fig:b_frag}:
\begin{enumerate}
\item The perturbative QCD calculation of the hard parton parton
  scattering cross-section convoluted with the parton densities and
  the strong coupling for the reactions $e p \to b \bar{b}$, $e p \to
  b \bar{b} g$ and $e p \to b \bar{b} q$
\item The fragmentation of the \bquark into a $B$-meson described by a
  non-perturbative fragmentation function
\item The semi-leptonic decay of the \bmeson into a muon and hadrons.
\end{enumerate}

\subsection{Perturbative LO and NLO QCD cross-section calculations}

The NLO QCD heavy quark cross-sections for DIS can be calculated using
the HVQDIS program~\cite{pr:d57:2806,np:b452:109,pl:b353:535}.  For
each event all contributions of the reactions $e p \to b \bar{b}$, $e
p \to b \bar{b} q$ and $e p \to b \bar{b} g$ are evaluated.  Since the
four-momenta of the outgoing particles are made explicitly available,
the single and double differential distributions as well as
correlations among all outgoing particles can be studied. Furthermore
HVQDIS allows to easily apply experimental cuts.

The \bquark production cross-section in $ep$ scattering can be
expressed as a product of the strong coupling, the parton densities
and the hard scattering parton cross-section which can be expanded in
a series of perturbatively calculable coefficient functions including
the matrix elements and the phase space factors.  While the total $ep$
\bquark cross-section is a well defined physical quantity which can be
compared to experimental data, the individual terms depend on the used
calculation scheme.

The divergences of the coefficient functions calculated to the order
${\cal O}(\alpha_s^2)$ due to soft gluon emissions are compensated by
contributions from virtual gluon exchange using the subtraction
method.  The renormalisation is carried out in the modified minimal
subtraction scheme ($\overline{MS}$) for light quarks.  The
divergences from heavy quark loops are subtracted such that the heavy
quark mass only enters in the coefficient functions. In the energy
evolution of the strong coupling and in the evolution of the parton
densities equations only light quarks appear.  The
CTEQ5F4~\cite{cteq5} parameterisation of the parton densities is used
as default. In this parameterisation the number of active quark
flavours in the proton is fixed to four.  As default, the mass of the
\bquark is set to $m_b=4.75 \GeV$. The uncertainty on this value is
assumed to be $\pm 0.25 \GeV$.

Uncertainties in the perturbative calculations are due to the choice
of the renormalisation ($\mu_r^2$) and the factorisation ($\mu_f^2$)
scales and due to the uncertainties on the $b$-mass. In addition the
uncertainty on the parton densities, in particular the gluon density,
contributes.  The uncertainty on the parton density function is
evaluated using the ZEUS NLO QCD fit to recent measurements of the
proton structure function~\cite{Chekanov:2002pv}.

\subsection{Fragmentation of the {\boldmath \bquark}}
\label{sec:b_frag}

The fragmentation of the \bquark into a \bmeson is a non-perturbative
process.  It can be described by a fragmentation function obtained
from comparisons with data.  The fragmentation of \bquarks has
recently been carefully studied in $e^+ e^-$ collisions using weakly
decaying $B$-mesons~\cite{Buskulic:1995gp}--\cite{Abe:2002iq}.
Several fragmentation functions describing the distribution of the
\bquark energy fraction carried by\pagebreak[3] the \bmeson have been tested and
parameterisations which describe the data best have been provided.
Since in $e^+ e^-$ collisions the initial energy of the \bquark is
known, they are usually parameterised as a function of the ratio of
the \bmeson energy ($E_B$) and the beam energy ($E_{\rm beam}$) which
corresponds to the energy of the \bquark:
\begin{equation}
x_B = \frac{E_B}{E_{\rm beam}} = \frac{2 E_B}{\sqrt{s}}\,,
\end{equation}
where $\sqrt{s}$ is the centre-of-mass energy of the collision.  The
variable $x_B$ can be interpreted as the fraction of the \bquark
energy which is carried by the \bmeson.

The \bquark fragmentation functions as a function of $x_b$ measured by
the SLD, OPAL and ALEPH collaborations are shown in
figure~\ref{fig:fragfun} together with some selected
parameterisations~\cite{Abe:2002iq,pr:d27:105,Kartvelishvili:1978pi}.
The data have been measured at a centre-of-mass energy which
corresponds to the mass of the $Z^0$-boson. They are consistent with
each other.  The parameterisations based on phenomenological models
are roughly able to reproduce the data.

{\renewcommand\belowcaptionskip{-1em}
\FIGURE[t]{\epsfig{file=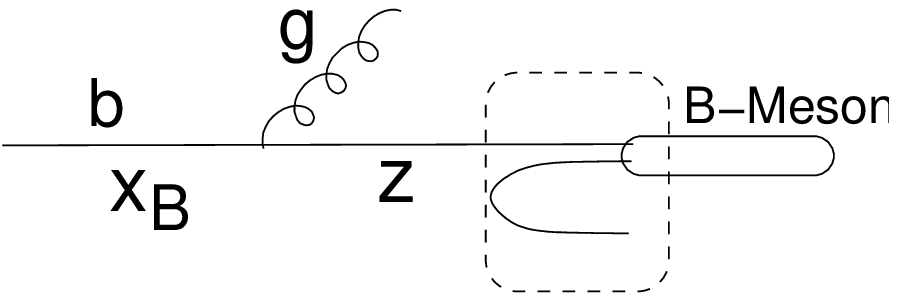,width=7.13cm}
  \epsfig{file=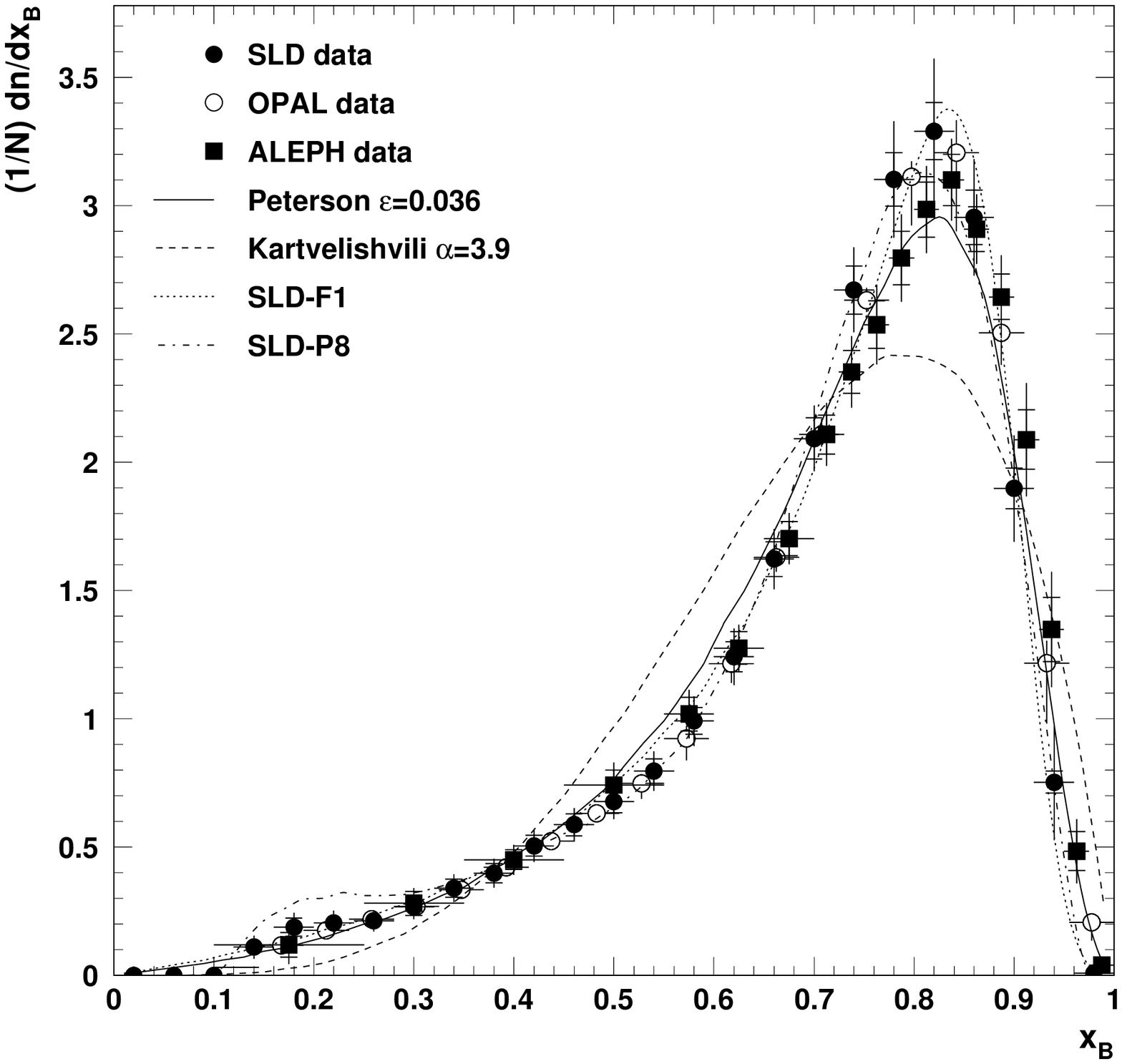,width=8.17cm}%
\caption{Shape of the distribution of the energy fraction $x_B= 2
  E_B/\sqrt{s}$ carried by the \bmeson with respect to the energy of
  the \bquark measured in $e^+ e^-$ collisions.  Shown are data
  together with some selected fragmentation function parameterisations
  (see text).\label{fig:fragfun}}}
}

However, the energy fraction of the \bquark just before entering the
fragmentation process, i.e.\ the \bquark energy fraction after
additional parton emissions, is not experimentally accessible (see
also the sketch on top of figure~\ref{fig:fragfun}).  The knowledge of
the distribution of this variable, called $z$ in the following, is,
however, necessary to make predictions for \bmeson production
cross-sections in other reactions.\footnote{In HVQDIS $z$ is defined
  as $\vec{p}_B = z \vec{p}_b$, where $\vec{p}_b$ ($\vec{p}_B$) is the
  momentum of the \bquark ($B$-meson).}

\pagebreak[3]

The fragmentation function $f(z)$ can in principle be obtained from
the \bmeson cross-section by folding them with the \bquark production
cross-section ($\sigma_b$):
\begin{equation}
\sigma_B \propto \int \sigma_b(z,\mu_{\rm FF}^2,\ldots ) \; 
f(z,\mu_{\rm FF}^2) dz\,,
\end{equation}
where $\mu_{\rm FF}^2$ is the factorisation scale which determines
which contributions should be absorbed in the perturbative part and
which in the non-perturbative fragmentation process.  The separation
between the pQCD calculation and the fragmentation function is
arbitrary, parts of the calculable contributions can be absorbed in
the fragmentation function and vice versa.  When evaluating the
physically observable \bmeson cross-section the perturbative and the
non-perturbative part have to be carefully matched to each other.  In
this sense the extracted fragmentation function always depends on the
scheme how the perturbative cross-section $\sigma_b$ has been
calculated.  As a consequence, when applying the fragmentation
function $f(z)$ for cross-section predictions in a different reaction,
care has to be taken that the same approximations are applied in the
perturbative calculation.  In praxis, this can be quite tricky, since
in a different reaction different parts of the perturbative
calculation, like e.g.\ resummations of logarithms etc., might have to
be taken into account.  Since in $e^+ e^-$ collisions at LEP or at the
SLAC linear collider the centre-of-mass energy is usually much larger
that the \bquark mass, in the perturbative calculable part large
logarithms of the form $\log{(s/m_b^2)}$ have to resummed
(NLL)~\cite{Mele:1991cw}.  It is not clear that such contributions
also have to be included for the cross-section calculations in
$p\bar{p}$-collisions or DIS.  Here, rather a resummation of terms
$\log{(E_{T,b}^2/m_b^2)}$, where $E_{T,b}$ is the \bquark transverse
energy, is needed~\cite{Cacciari:1998it,Cacciari:1994mq}.  

As default, we follow the suggestion in
refs.~\cite{Cacciari:2002pa,Cacciari:2002gn} to use fragmentation
functions where the fourth Mellin moment, defined as $f_N = \int_0^1
z^{N-1} f(z) dz$ with $N=4$, folded with a perturbative NLL
calculation is in agreement with the $e^+ e^-$ data.  Such a procedure
is optimal when the calculations are compared to data which exhibit a
cross-section behaviour like $d\sigma_b/dE_{T,b} \propto
1/dE_{T,b}^4$, which is the case in \bquark production at Tevatron as
well as in DIS at HERA for large transverse momenta of the \bquark
(see line in figure~\ref{fig:bquark}).\footnote{The reason is that
  then in the moment space the \bquark cross-section is just given as
  product of $1/E_{T,b}^n$ and the $n$th-moment of the fragmentation
  function $f_N(z)$, since\\ $d\sigma_b/dE_{T,b} \propto \int dz
  d\hat{E}_{T,b} \, f(z) \, \frac{1}{\hat{E}_{T,b}^N} \, \delta(z
  \hat{E}_{T,b} - E_{T,b}) = \frac{1}{E_{T,b}^N} \, f_N(z).$} The
parameterisation of Kartvelishvili et
al.~\cite{Kartvelishvili:1978pi}, $f(x,\alpha) = (\alpha + 1) \,
(\alpha + 2) \, x^\alpha \, (1 - x)$ is
proposed~\cite{Cacciari:2002pa,Cacciari:2002gn}, since it is
consistent with an expansion in powers of $\lambda/m_b$, where
$\lambda \approx 0.4$ is a typical hadronic scale, as it is predicted
by QCD~\cite{Jaffe:1994ie,Nason:1997pk,Cacciari:2002xb}.  The
parameter\footnote{Note, that this value for $\alpha$ is quite
  different then the one best describing the $(1/N) dn/dx_b$
  distributions, namely $\alpha=3.9$~\cite{Abe:2002iq}, which is shown
  in figure~\ref{fig:fragfun}.} $\alpha = 27.5$ is obtained from a fit
to the ALEPH
data~\cite{Cacciari:2002pa,Heister:2001jg,Cacciari:2002gn}.  The
perturbative part has been calculated to NLO accuracy together with a
resummation of logarithmic terms to all orders. It is not clear
whether such a resummation is also needed for the range of $E_{T,b}$
presently accessible in DIS at HERA.

Therefore and also since the spectrum of the transverse \bquark
momentum deviates from the power-law behaviour (see above) at low
values, one can use different parameterisations with different
adjustable parameters to investigate the uncertainty due to the
fragmentation function.  As alternative e.g.\ the
parameterisation\footnote{The explicit form is: $N \frac{1}{z} (1 -
  \frac{1}{z} - \frac{\epsilon}{1 -z})^{-2}$, where $N$ is a
  normalisation factor and $\epsilon$ is an adjustable parameter,
  which should be in the order of $\epsilon = m^2/M_B^2 \approx
  0.002$, where $m$ ($M_B$) is a typical mass of a light (B) meson.
  Note, that the value $\epsilon = 0.002$ is a crude estimation and
  not a result of a fit to the data.}  of Peterson et
al.~\cite{pr:d27:105} can be used.

\subsection{The semi-leptonic decay of the {\boldmath \bmeson} into a muon and a jet}\label{sec:B_decay}

After the \bmeson is produced it decays via a charged current
interaction.  In some cases the \bmeson decays semi-leptonically,
e.g.\ a muon and a jet is produced. The muon can either be produced by
a direct semi-leptonic decay ($B \to \mu X$) or by an indirect decay
where the \bquark decays first to a charm quark which subsequently
decays semi-leptonically to a muon ($B \to c X \to \mu X$).  About $10
\%$ of all \bmeson decays are direct decays into muons, in $8 \%$
($2\%$) of the cases the muons are indirectly produced by a charm
(anti-charm) decay.  Other decay modes of the \bquark are much
smaller, e.g.\ $B \to J/\psi X \to \mu X$ is approximately $7 \cdot
10^{-4}$ and $B \to \tau X \to \mu X$ is $\approx 7 \cdot 10^{-3}$ .
The branching fractions presently implemented in JETSET agree within
$3 \%$ with the measured ones.  The agreement in the different
channels is reasonable.  In the HVQDIS calculation the cross-section
is determined using the sum of the branching ratios of direct and
indirect decays of the \bmeson\footnote{The term $B$-meson refers here
  to the mixture used in JETSET, i.e.\ $0.4025 \, B^0 + 0.4025 \,
  B^+ + 0.094 \, B_s + 0.101 \, \Lambda_b$.  This mixture is in
  agreement with the data~\cite{Hagiwara:2002fs}:  $(0.38 \pm 0.13)
  \, B^0 + (0.388 \pm 0.13) \, B^+ + (0.106 \pm 0.13) \, B_s + (0.118
  \pm 0.2) \, \Lambda_b$.  }  decays into a muon is ${\cal B}_\mu =
0.22$.

In the calculation of the cross-section $ e p \to b \bar{b} \to \mu X$
it has to be taken into account that the measured muon can stem from
the \bquark or the $\bar{b}$-quark. For this purpose, it is assumed
that the \bquark and the $\bar{b}$-quark are both independently
decaying into muons according to the branching ratio ${\cal
  B}_\mu$. We ignore the case where a \bquark produces two muons via
$b \to c \mu X \to \mu \mu X$.  If one of the muons is in the detector
acceptance, the event is counted in the calculation.
To improve the
efficiency of the calculation, one can as an alternative always decay
one of the \bquarks in a muon and then decay the other one according
to the probability that exactly two muons are in the event under the
condition that there is at least one muon, i.e.\ $ P_{2 \mu | \ge 1
  \mu} = {\cal B}_\mu^2/(2 {\cal B}_\mu - {\cal B}_\mu^2) = {\cal
  B}_\mu/(2 - {\cal B}_\mu)$.  The cross-section has then to be
multiplied with the probability that there is at least one muon
$P_{\ge 1\mu} = 1 - (1 - {\cal B}_\mu) (1- {\cal B}_\mu) = 2 
   {\cal B}_\mu - {\cal B}_\mu^2$.  To ensure the cancellation of the different
divergences appearing in the NLO QCD calculation care has to be taken
that for one event defined, e.g.\ by a common $x$ and $Q^2$ value,
only one common $z$-value is used for the different contributions
appearing in a NLO QCD calculation.

\FIGURE[t]{\epsfig{file=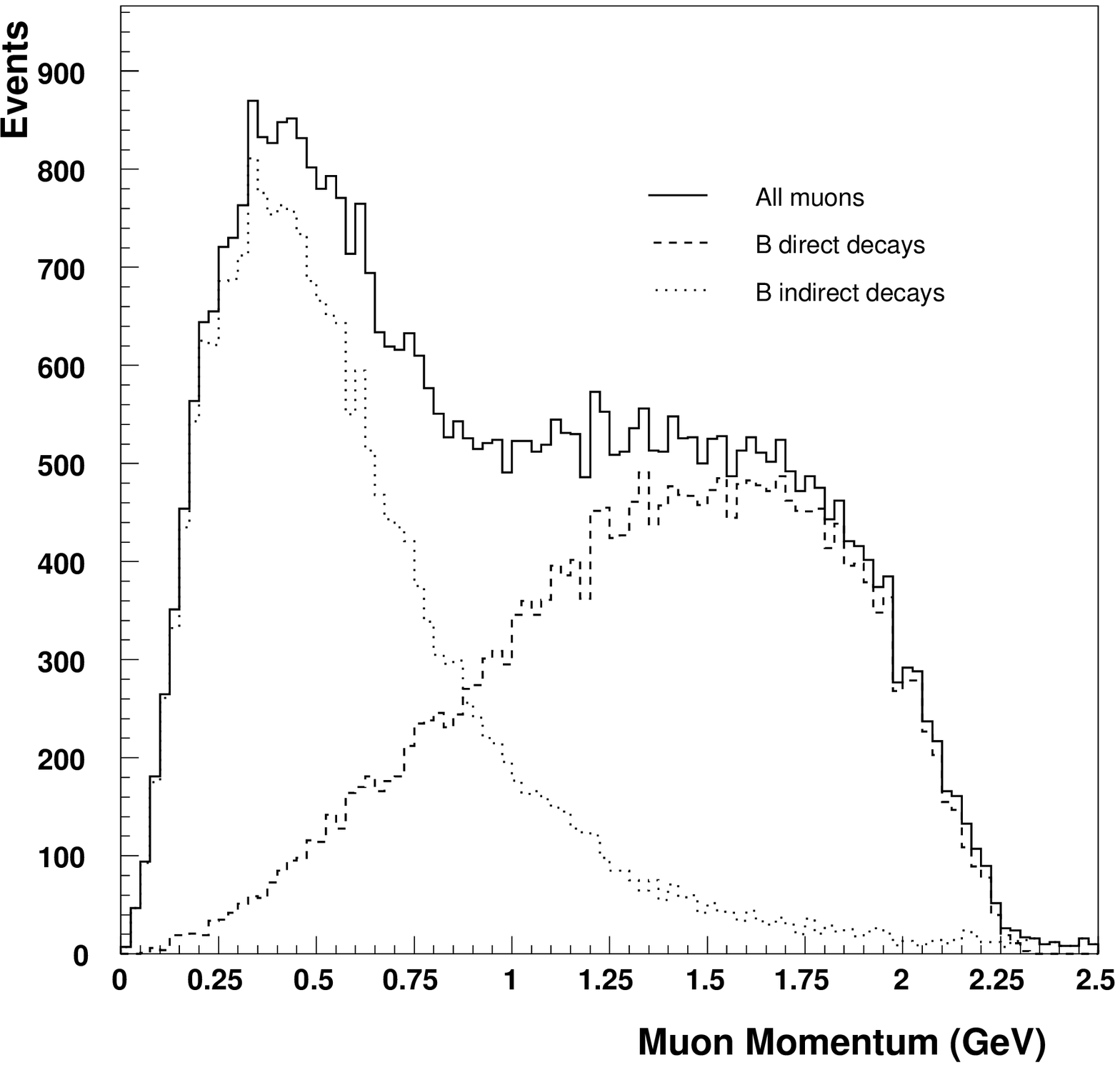, width=.7\textwidth}%
\caption{Momentum distribution of the decay muons in the rest frame of
  the parent \bmesonx. The contributions from direct ($b \to \mu X $)
  and indirect ($b \to c X \to \mu X$) decays are shown separately.
\label{fig:slspectrum}}}

To correctly model the kinematics of the muon decay a parameterisation
of the muon momentum spectrum in the centre-of-mass frame of the
\bmeson is needed. Since the muon spectrum for direct $(B \to \mu X)$
and indirect $(B \to c X \to \mu X)$ \bmeson decays is different, the
correct mixture has to be used.  Muons are generated isotropically in
the rest frame of the \bmesonx.  The absolute momentum $p$ is obtained
with the help of a probability function extracted from the muon
momentum spectrum as obtained from JETSET.
Figure~\ref{fig:slspectrum} shows the muon momentum distribution in
the rest frame of the \bmesonx, $dN/dp$.  The total distribution is
shown as solid line, the distributions corresponding to the direct and
indirect $B$-decays are shown as dashed and dotted lines separately.
The direct \bmeson decays produce events with higher muon momenta.

\section{Comparison of the leading order calculations}\label{sec:lo_calc}

To check that the simple form of \bquark fragmentation and of the
subsequent decay of the \bmeson implemented in HVQDIS gives reasonable
results, a comparison to the more complex fragmentation models used in
the Monte Carlo simulation programs can be made.  For this purpose,
the same free parameters in the perturbative cross-section
calculations have to be chosen. A comparison of the LO QCD prediction
of the inclusive DIS \bquark cross-section calculated by HVQDIS and by
RAPGAP\footnote{To obtain a pure LO calculation in RAPGAP, the parton
  shower option has been switched off. Moreover, for the calculation
  of the differential \bquark and jet cross-section the fragmentation
  has been switched off.} is shown in figure~\ref{fig:vgllomeps}.  The
renormalisation and factorisation scale $\mu^2 = \mu_f^2 = \mu_r^2 =
Q^2$ has been used and the \bquark mass was set to $m_b = 5 \GeV$.
Moreover the $1$-loop formula of the strong coupling $\alpha_s$ and
the CTEQ5L parton density functions have been used.

{\renewcommand\belowcaptionskip{-1.3em}
\FIGURE[t]{\begin{tabular}{l}
\epsfig{file=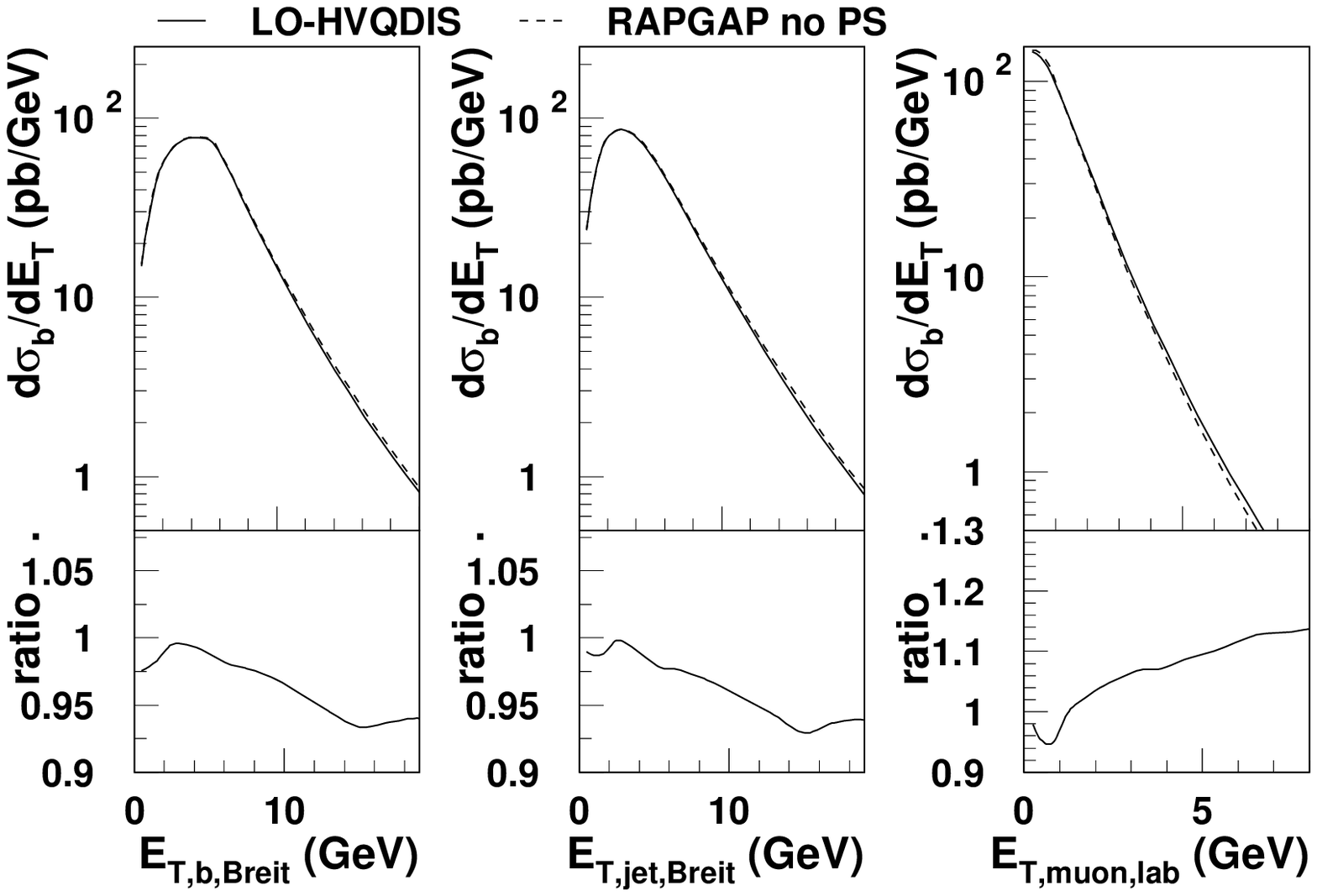,width=.6\textwidth}\\[-15pt]
\small\hspace{5pt}($a$)\hspace{73pt}($b$)\hspace{75pt}($c$)\\[5pt]
        \epsfig{file=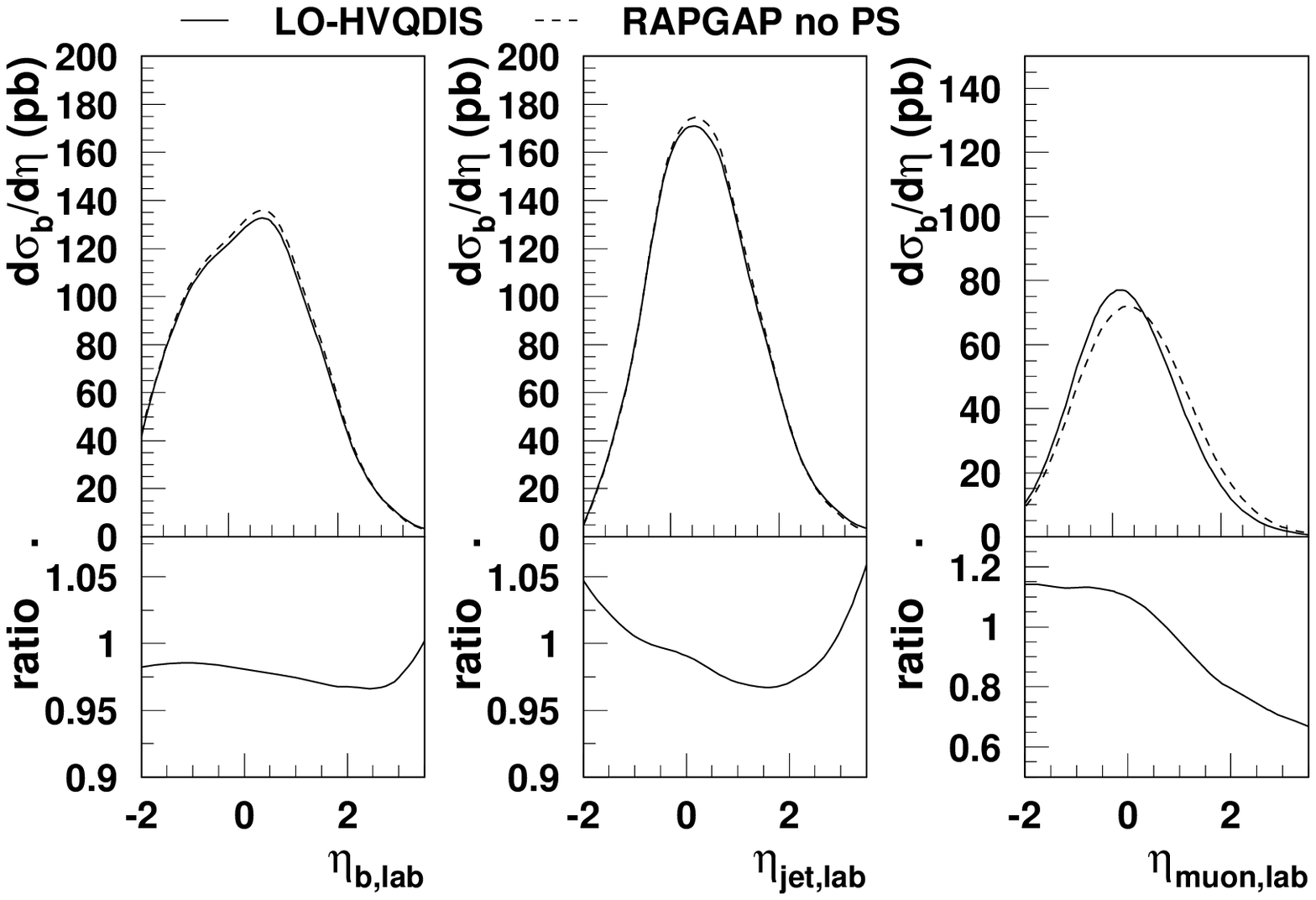,width=.6\textwidth}\\[-15pt]
\small\hspace{5pt}($d$)\hspace{73pt}($e$)\hspace{75pt}($f$)%
	\end{tabular}
\caption{Inclusive DIS \bquark cross-sections as a function of 
($a$) the transverse energy of the \bquark \etb,
($b$) the transverse energy of the jet \etjet,
($c$) the transverse energy of the muon \etmu,
($e$) the pseudo-rapidity of the jet \etajet,
($d$) the pseudo-rapidity of the \bquark \etab,
($f$) the pseudo-rapidity of the muon \etamu.
Shown are LO QCD cross-section calculated by HVQDIS (LO) and by
RAPGAP without parton showers. More details are given in the text.
\label{fig:vgllomeps}}}
}

The inclusive DIS \bquark cross-section as calculated by LO-HVQDIS is
$\sigmab = 530\pb$ \linebreak and agrees within $2\%$ with the
prediction from RAPGAP.  For the \bquark cross-section requiring a jet
$\sigmabj = 148\pb$ is found.  The agreement with RAPGAP is within
$4\%$.  This is a strong indication that the LO matrix elements used
in the two programs are the same. However, difference up to $5\%$ are
found for large transverse energies of the \bquark \etb~and of the jet
\etjet.  This can be seen in figure~\ref{fig:vgllomeps} where the
inclusive \bquark cross-section \sigmab~as a function of the
transverse energy of the \bquark \etb~(a) and as a function of the
transverse jet energy \etjet~(b) in the Breit frame is shown. No
strong dependence on the pseudo-rapidities of the \bquark \etab~or of
the jet pseudo-rapidities in the laboratory frame \etajet~is found
(see figures~\ref{fig:vgllomeps}$d$ and~\ref{fig:vgllomeps}$e$).

For the \bquark cross-section requiring a jet and a muon $\sigmabjm =
19\pb$ is found with LO-HVQDIS, while the one from RAPGAP is $10\%$
larger.  The observed difference in $\sigmabjm$ is probably due to an
incomplete\pagebreak[3] modeling of the muon fragmentation in the HVQDIS program.
This can be seen in figures~\ref{fig:vgllomeps}$c$
and~\ref{fig:vgllomeps}$f$.  The transverse energy spectrum of the muon
is harder for the HVQDIS calculation. At large transverse energies it
is $10\%$ higher. The biggest difference is found in the
pseudo-rapidity spectrum. In the RAPGAP calculation the muon is
produced more forward. Difference up to $30 \%$ are found.  A possible
explanation of this effect is that the fragmentation model in HVQDIS
is too simple to cope with the multi-parton environment where it is
important that all colour connections are correctly defined.

\vspace{-1pt minus 10pt}

\section{Perturbative NLO QCD calculations and their uncertainties}\label{sec:uncertain}
\vspace{-1pt minus 6pt}

\subsection{Calculated NLO QCD cross-sections}
\vspace{-1pt minus 4pt}

\TABLE[b]{\begin{tabular}{|c|cccc|}
\hline NLO & \sigmab ~({\rm pb}) & \sigmabj
  ~({\rm pb}) & \sigmabm ~({\rm pb}) & \sigmabjm ~({\rm pb}) \\ \hline \bquark &
  $598$ & $230$ & $53$ & $ 26$ \\ $c$-quark & $24170$ & $2320$ & $515$
  & $188$\\\hline
\end{tabular}
\caption{NLO QCD cross-sections calculated for the four studied
  cross-section definitions for \bquarks and $c$-quarks.
\label{tab:cross-section_nlo}}}

The NLO QCD \bquark cross-sections for the four cases \sigmab,
\sigmabj, \sigmabm, \sigmabjm~are given in
table~\ref{tab:cross-section_nlo}.  The cross-sections have been
calculated with the renormalisation and factorisations scale set to
$\mu^2 = p_{T,b}^2 + 4 m_b^2$, where $p_{T,b}$ is the transverse
momentum of the \bquark in the Breit or hadronic centre of mass
frame.\footnote{The hadronic centre of mass frame is defined as
  $\vec{q} + \vec{P} = 0$, where $q$ ($P$) is the photon (proton)
  momentum.  The centre of mass frame can be transformed by a
  longitudinal boost to the Breit frame.  The transverse energy is
  therefore the same in both frames.}  The \bquark mass has been set
to $m_b= 4.75$\GeV. The CTEQ5F4 parameterisation has been used for
the parton densities.  Also shown are the charm quark
cross-sections.\footnote{The charm quark cross-section have been
  calculated with the factorisation and renormalisation scales set to
  $\mu^2 = p_{T,c}^2 + 4 m_c^2$ and the charm quark mass is set to
  $m_c = 1.4 \GeV$. The CTEQ5F3 parton density functions and the
  Peterson fragmentation function with $\epsilon = 0.033$ have been
  used. The semi-leptonic muon momentum distribution and the branching
  fraction ${\cal B}_{c \to \mu X}$ have been modified for charm
  quarks.} 
 
The inclusive charm quark cross-section is about $370$ times larger
than the inclusive \bquark cross-section. This is also expected, since
the electrical charge of the charm quark is higher and its mass is
smaller.  The heavy quark mass enters directly in the matrix element
and in the phase space factor, in addition it changes the parton
kinematics such that different parton densities are probed.  It is,
however, interesting that if a hard jet or a muon is required in the
detector acceptance the charm quark cross-section is only an order of
magnitude bigger than the \bquark cross-section. The reason for this
behaviour is that the large \bquark mass naturally produces particles
at higher transverse energy.

The measurable \bquark cross-section (\sigmabjm) is about $20$ times
smaller than the inclusive one (\sigmab).  Only part of this
difference is due to the branching fraction.  The large fraction of
the \bquark cross-section can not be measured in the detector, since
the \bquark has either a low transverse energy or is produced in the
forward direction\footnote{At HERA, the proton moves into the
  $+z$-direction. The forward direction is the therefore the region
  located toward the proton remnant.}  outside the detector
acceptance.

\pagebreak[3]

The \bquark cross-section as a function of the \bquark transverse
energy in the Breit frame and its pseudo-rapidity in the laboratory
frame is shown in figure~\ref{fig:bquark}.  The central NLO QCD
predictions are shown as lines.  The inner band indicates the
renormalisation and factorisation scale uncertainties, the outer bands
indicates in addition the uncertainty due to the \bquark mass.  For
the inclusive \bquark cross-section (\sigmab) the \bquark transverse
energy distribution exhibits a broad peak at about $E_{T,b,Breit} = 5$
\GeV. Above this value the distribution falls like
$d\sigmab/dE_{T,b,Breit} \propto E_{T,b,Breit}^{-4}$ toward larger
values. Below this value the \bquark cross-section is approximately
constant.  If a hard jet is required (dashed or dashed-dotted line in
figure~\ref{fig:bquark}), the behaviour at large $E_{T,b,Breit}$ is
not modified, but the region of $E_{T,b,Breit}$ is suppressed. When
measuring the \bquark cross-section via the semi-leptonic muon decay,
the NLO QCD predictions for relatively large $E_{T,b,Breit}$ are
tested.  For all shown cross-sections the uncertainties at low
transverse \bquark energies are larger.  Figure~\ref{fig:bquark}$b$
shows the \bquark pseudo-rapidity in the laboratory frame.  For all
cross-section definitions most of the \bquarks are produced centrally,
i.e.\ $-2 < \eta_{b,lab} < 2$. However, in the inclusive case
(\sigmab) about $10\%$ of the \bquarks are produced at
pseudo-rapdities beyond $\eta_{b,lab} > 2.5$, where there is no
detector coverage. If the muon is required to be in the detector
acceptance, only \bquarks within $-2 < \eta_{b,lab} < 2$ contribute to
the cross-section.

{\renewcommand\belowcaptionskip{1em}
\FIGURE[t]{\epsfig{file=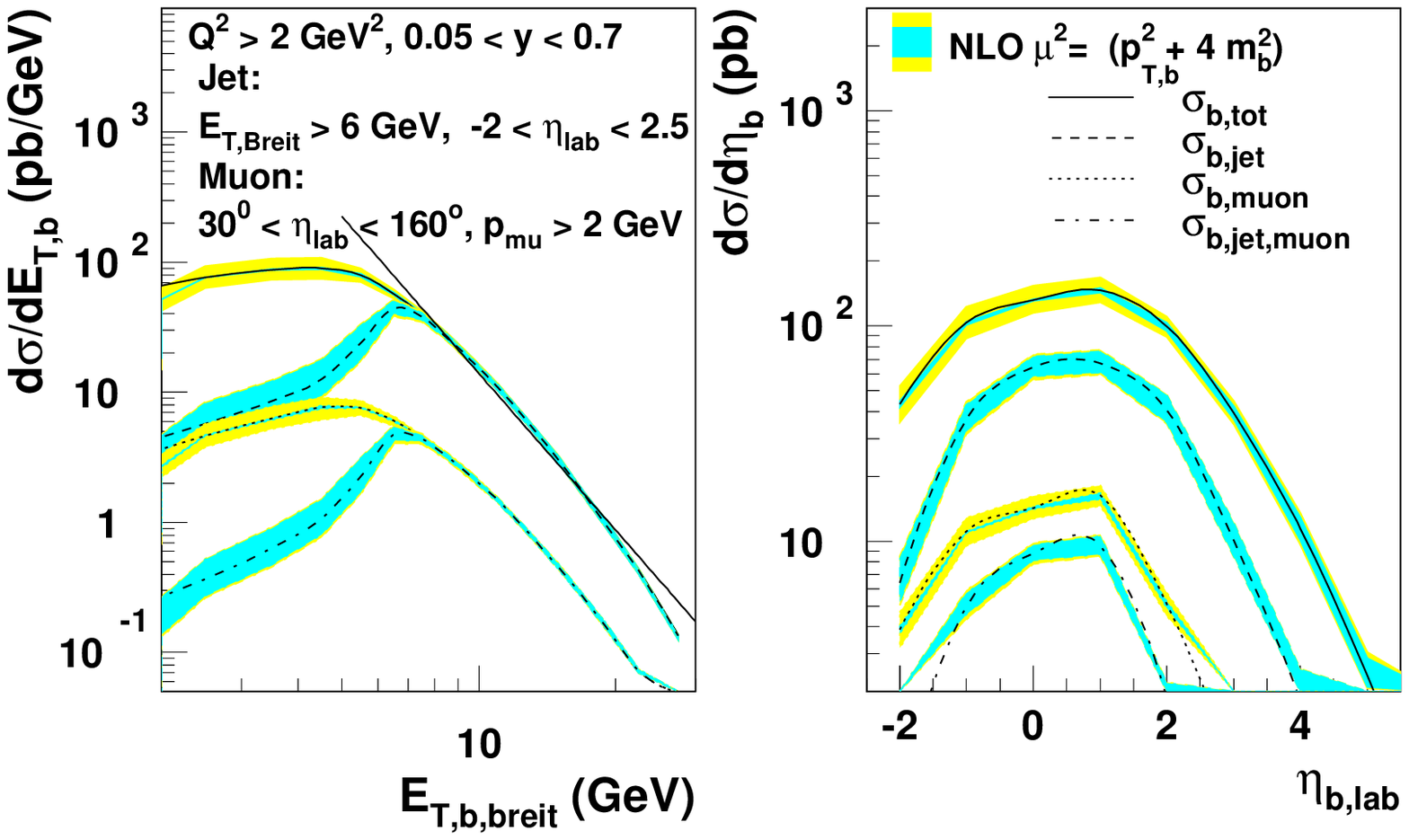,width=.75\textwidth}%
\begin{picture}(0,0) 
\put(-290,0){($a$)}
\put(-130,0){($b$)} 
\end{picture} 
\caption{Cross-sections (\sigmab, \sigmabj, \sigmabm, \sigmabjm) as a
  function of the transverse energy ($a$) and the pseudo-rapidity of
  the \bquark for the four studied definition (see text).  The inner
  band represents the renormalisation and factorisation scale
  uncertainties and the outer band in addition the uncertainty due to
  the \bquark mass.  The renormalisation and factorisation scale has
  been set to $\mu^2 = p_{T,b}^2 + 4 m_b^2$.  As a comparison, the
  solid line indicates the power law $\sigma_b/dE_{T,b} \propto
  E_{T,b}^{-4}$.
\label{fig:bquark}}}}

\subsection{Factorisation and renormalisation scale and {\boldmath \bquark} mass uncertainties}

A fixed order QCD cross-section calculation usually
depends on the renormalisation and the factorisation scales. The
renormalisation procedure removes divergences due to loop
contributions in which virtual exchanged partons can have very large
momenta (ultra-violet divergences). The factorisation scale separates
the perturbatively calculable short distance contributions from the
non-perturbative long-distance contributions. The non-perturbative
part is absorbed in the parton density functions.  In this way
divergences due to collinear parton radiation are removed.  The choice
of the renormalisation and factorisation scales in the calculation is
arbitrary, but should correspond to the hard scale (${\cal Q}^2$)
involved in the hard scattering process.  In a complete calculation a
physical observable does not depend on the choice of the
renormalisation and factorisation scale. The residual scale dependence
in a fixed order calculation can be used to estimate the uncertainties
due to neglected contributions.

In a LO calculation the calculated cross-section depends on the
renormalisation scale via the energy behaviour of the strong coupling
$\alpha_s(\mu_r)$. When the renormalisation scale is increased, it is
therefore expected that the cross-section decreases in the same way.
In a NLO calculation the $\mu_r$ dependence of $\alpha_s$ can be
compensated by the perturbatively calculable coefficient functions,
since they also depend on $\mu_r$. The scale dependence can therefore
be significantly reduced when the NLO corrections are included in the
calculation.  In some cases the residual scale dependence is closely
related to the higher order contributions which have not been
calculated.

\looseness=-1 The dependence on the scale factor $\xi$, defined as
$\mu^2 = \xi{\cal Q}^2$, where ${\cal Q}$ is the hard scale, is
shown in figure~\ref{fig:dsigvsmu}$a$ for the four different
cross-section definitions as calculated in NLO QCD. The hard scale has
been chosen to be ${\cal Q}^2 = Q^2 {+} 4 m_b^2$.  The cross-sections are
normalised to the cross-sections obtained for $\xi = \xi_0 = 1$, i.e.\
the ratio $(d\sigma/d\xi)/(d\sigma/d\xi_0)$ is shown.  When $\xi$ is
varied by about a factor of $10$ in both directions, the inclusive
cross-section \sigmab~varies by about $\pm 5 \%$. Once a hard jet is
required, the cross-section variation increases to $\pm 20 \%$.

\FIGURE[t]{\epsfig{file=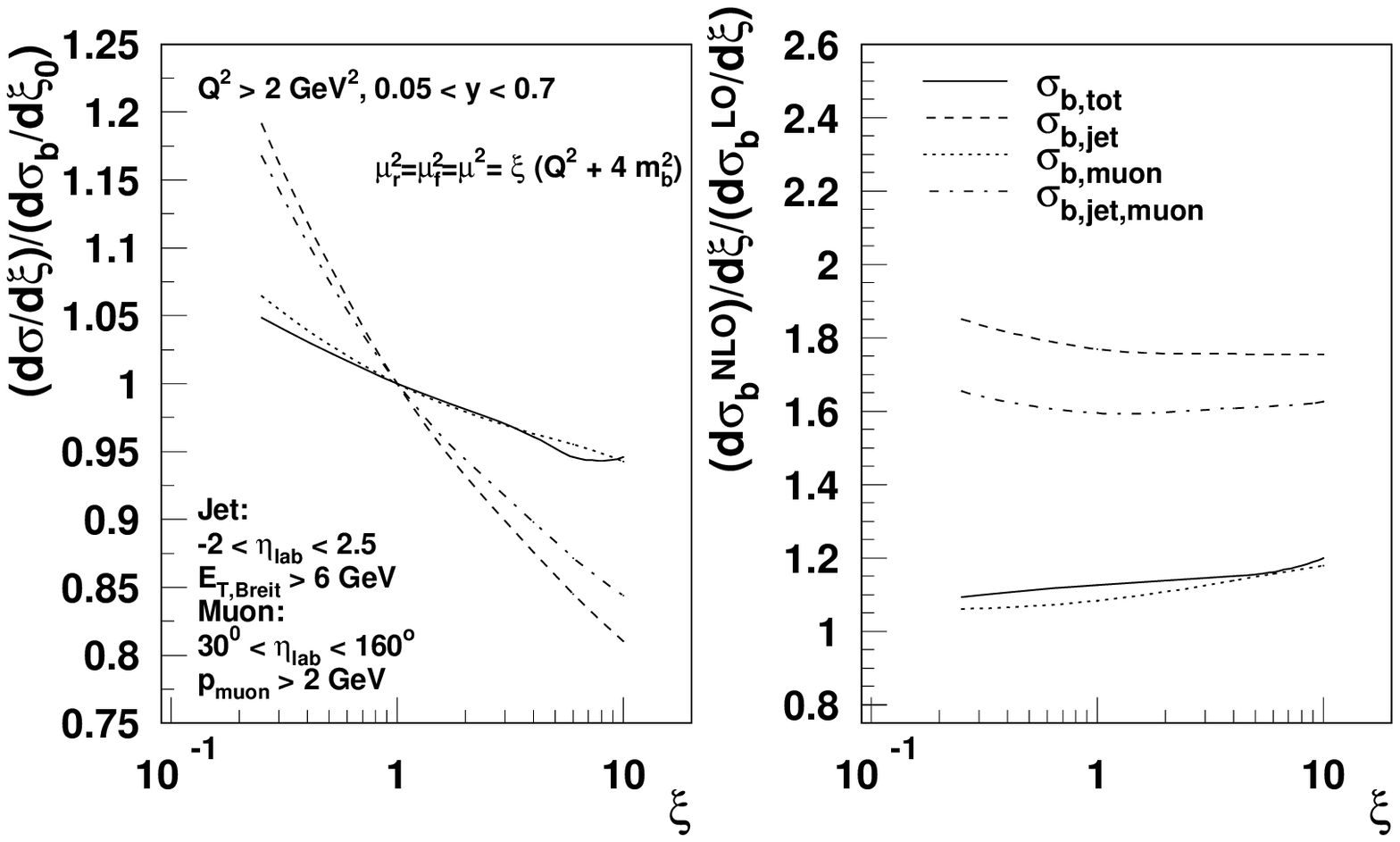,width=.85\textwidth}%
\begin{picture}(0,0) 
\put(-310,-5){($a$)}
\put(-150,-5){($b$)} 
\end{picture} 
\caption{($a$) Ratio of the cross-sections \sigmab, \sigmabj,
  \sigmabm, \sigmabjm~ calculated with the renormalisation and
  factorisation scales set to $\mu^2 = \xi (Q^2+4 m_b^2)$ and $\mu^2 =
  \xi_0 (Q^2+4 m_b^2)$ where $\xi_0=1$.  ($b$) Ratio of the
  cross-sections calculated at NLO and at LO (K-factor) as a function
  of the scale factor $\xi$.\label{fig:dsigvsmu}}}

Figure~\ref{fig:dsigvsmu}$b$ shows the NLO correction as a function of
the scale factor $\xi$.  The NLO correction is defined as the ratio of
the cross-section calculated at NLO to the one calculated at LO.  The
dependence of the NLO correction on the scale factor is rather small
for all studied cross-section definitions. While the NLO correction is
only about $10\%$ for the inclusive cross-section \sigmab~and for the
one requiring a muon \sigmabm, it increases to almost a factor of $2$
when a jet is required.

This together with the stronger scale dependence is an indication that
for the cross-section measured via the $P_T^{rel}$-method where a jet
has to be required higher order contributions are more
important. However, whenever a hard jet is involved in the
cross-section measurement, the increased uncertainty is unavoidable.
It is interesting that the requirement of a muon in the detector
acceptance, does not lead to an increased uncertainty.  This is
clearly an advantage for cross-section measurements based on the
\bmeson lifetime using displaced vertices.  However, also in this case
the requirement of a hard jet in the Breit frame might be necessary in
order to efficiently reject the large background from inclusive DIS
events.

\looseness=1Here and in the following the renormalisation and factorisation scales
have been set equal. Since this is an assumption with no deeper
justification we have, for ${\cal Q}^2 = p_{T,b}^2 + 4 m_b^2$, also
varied $\mu_r^2$ and $\mu_f^2$ independently. For all factorisation
(renormalisation) scales the cross-section rises for decreasing
(increasing) $\mu^2_r$ ($\mu^2_f$).  The \bquark cross-section changes
from $\sigmab= 597.6 \pb$ for $\mu^2_f= \mu_r^2= p_{T,b}^2 + 4 m_b^2$
to $\sigmab= 671.3 \pb$ for $\mu^2_f= 4 \, (p_{T,b}^2 + 4 m_b^2)$ and
$\mu^2_r= 0.25 \, (p_{T,b}^2 + 4 m_b^2)$ and to $\sigmab= 511 \pb $
for $\mu^2_f= 0.25 \, (p_{T,b}^2 + 4 m_b^2)$ and $\mu^2_r= 4 \,
(p_{T,b}^2 + 4 m_b^2)$. Similar conclusion hold for the other
cross-section definitions.  We find that the variation with the
renormalisation scale is about equal for all choices of the
factorisation scale, i.e.\ about $\pm 5 \%$ for \sigmab~and about $\pm
20 \%$ for \sigmabjm. Also the dependence of the factorisation scale
is about equal for all choices of the renormalisation scale, i.e.\
about $\pm 5 \%$ for \sigmab~and also for \sigmabjm.  The fact that
the renormalisation scale dependence does not depend on the
factorisation scale choice and vice versa, remains valid even for
extreme scale factors like $\xi = 40$.  In the $\mu_f^2-\mu^2_r$ plane
we therefore have not found a plateau region where the scale
dependence is significantly reduced with respect to other regions.

\looseness=1Figure~\ref{fig:bcross_q24m2} shows the cross-sections \sigmab,
\sigmabj, \sigmabm, \sigmabjm~as a function of $Q^2$ and of
$E_{T,jet,Breit}$. From $2 < Q^2 < 1000 \GeVsq$ the cross-sections
fall by about $4$ orders of magnitudes toward larger $Q^2$
(figure~\ref{fig:bcross_q24m2}$a$).  The $Q^2$ dependence is similar for
all studied cross-section definitions.  At low $Q^2$,
\sigmabj~(\sigmabjm) is about a factor $3$ (2) smaller than the
inclusive cross-section \sigmab~(\sigmabm), at $Q^2 \approx 200
\GeVsq$ the ratio $\sigmab/\sigmabj$ ($\sigmabm/\sigmabjm$) is only
$2$ ($1$).  The dependence on the transverse jet energy is similar for
all cross-section definitions for $E_{T,jet,Breit} > 6 \GeV$ (see
figure~\ref{fig:bcross_q24m2}$b$).  The inclusive cross-section
\sigmab~as well as the one requiring a muon \sigmabm~ result in a mean
transverse jet energy of about $10$\GeV. When a hard jet is required
the mean transverse jet energy increases to about $15$\GeV. As a
consequence possible terms of the form
$\log{(E_{T,jet,Breit}^2/m_b^2})$ increase from about $0.5$ to $1$.
About $35\%$ of the inclusive \bquark events have a hard jet with
$E_{T,jet,Breit} > 6 \GeV$, when a muon in the detector acceptance is
required $45\%$ of the events have a hard jet.

{\renewcommand\belowcaptionskip{1em}
\FIGURE[t]{\epsfig{file=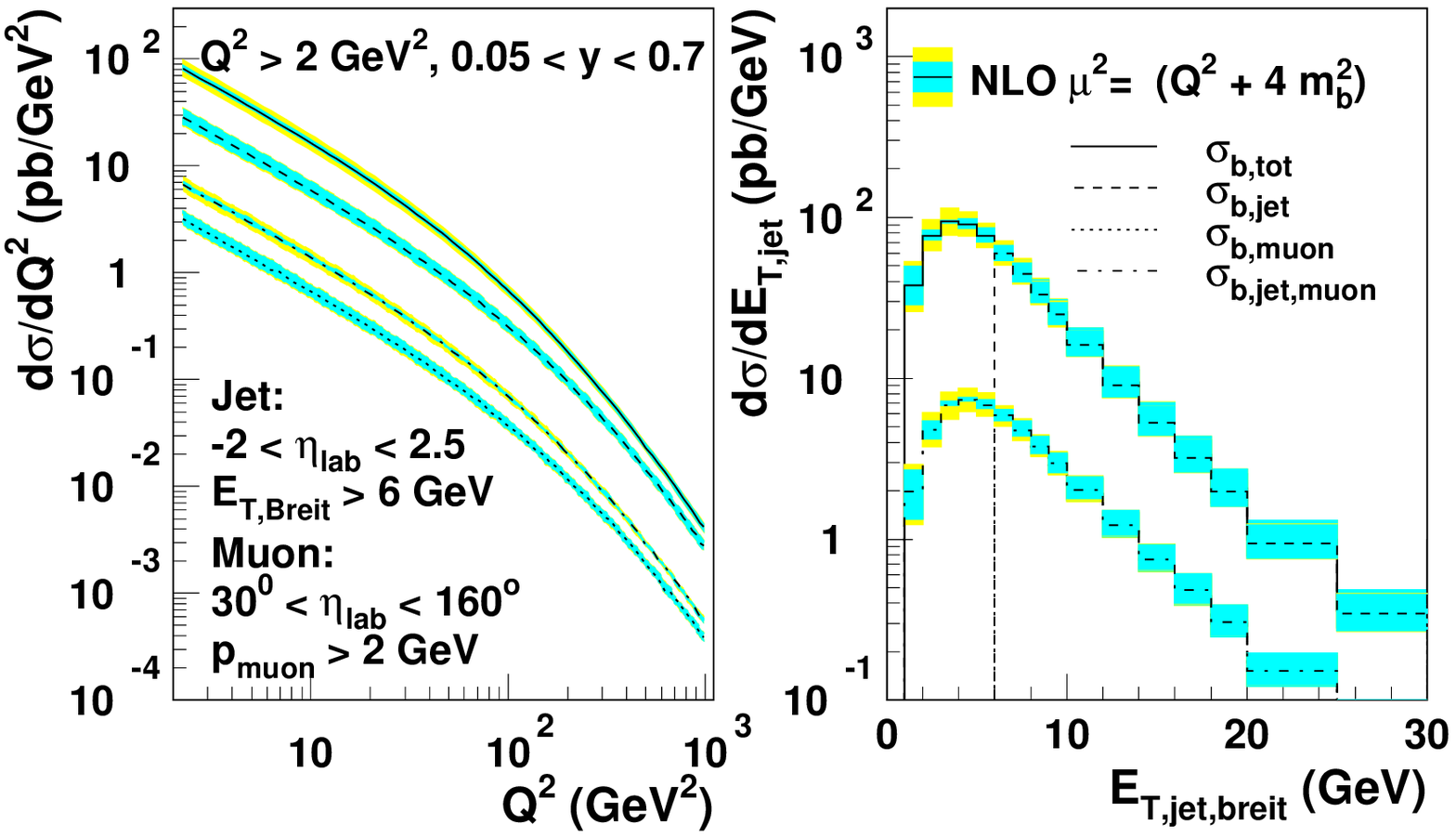,width=.75\textwidth}%
\begin{picture}(0,0) 
\put(-290,0){($a$)}
\put(-130,0){($b$)} 
\end{picture} 
\caption{Cross-sections (\sigmab, \sigmabj, \sigmabm, \sigmabjm) as a
  function of $Q^2$ ($a$) and the transverse jet energy \etjet ($b$).
  The inner band represents the renormalisation and factorisation
  scale uncertainties and the outer band in addition the uncertainty
  due to the \bquark mass.  The renormalisation scale has been set to
  $\mu_r^2 = Q^2 + 4 m_b^2$.\label{fig:bcross_q24m2}}}}

The uncertainty on the renormalisation and factorisation scales can be
estimated by varying the scale factor $\xi$.  It is common practice to
vary $\mu^2$ by approximately one order of magnitude, i.e.\ the scale
factor is varied by $0.25 < \xi < 4$.  In addition, the uncertainty
introduced by the limited knowledge on the \bquark mass can be
estimated by varying it between $4.5 < m_b < 5.0$\GeV.  The default
value is set to $m_b = 4.75 \GeV$.

{\renewcommand\belowcaptionskip{1em}
\FIGURE[t]{\epsfig{file=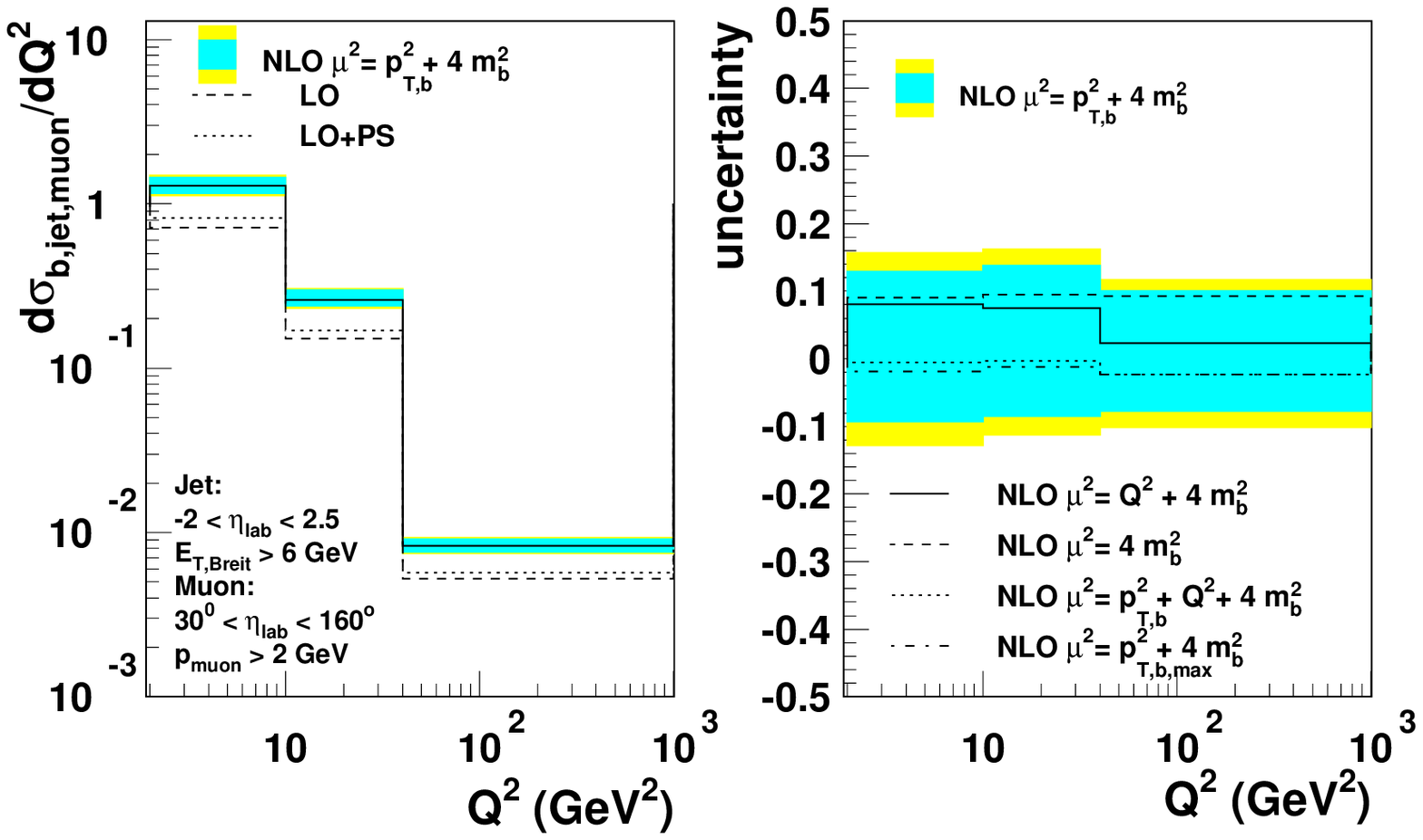,width=.75\textwidth}%
\begin{picture}(0,0) 
\put(-290,0){($a$)}
\put(-130,0){($b$)} 
\end{picture} 
\caption{($a$) Bottom quark cross-section \sigmabjm~as a function of
  $Q^2$ for different choices of the renormalisation and factorisation
  scales $\mu^2$.  ($b$) Scale uncertainty for the choice $\mu^2=
  p_{T,b}^2 + 4 m_b^2$ (band) and the ratio of the cross-sections when
  calculated with different scale choice normalised to the one
  calculated with $\mu^2= p_{T,b}^2 + 4 m_b^2$ (lines).  The inner
  band represents the renormalisation and factorisation scales
  uncertainties and the outer band in addition the uncertainty due to
  the \bquark mass.\label{fig:dsigdq2_renscaleall}}}}

Figure~\ref{fig:dsigdq2_renscaleall}$a$ shows the measurable \bquark
cross-section \sigmabjm~as a function of $Q^2$. The hard scale is set
to ${\cal Q}^2 = p_{T,b}^2 + 4 m_b^2$.  The inner band illustrates the
scale uncertainty, the outer band shows in addition the uncertainty
due to the \bquark mass. The size of the uncertainty is better visible
in figure~\ref{fig:dsigdq2_renscaleall}$b$ where the relative difference
of the modified cross-section calculation to the default cross-section
calculation, i.e.\ $(\sigmabjm(\xi_0 \pm \xi, m_b \pm \Delta m_b) -
\sigmabjm(\xi_0, m_b))/ \sigmabjm(\xi_0, m_b)$, is shown. The total
uncertainty is about $\pm 15\%$ at low $Q^2$ and slightly decreases
towards larger $Q^2$.

Also shown in figure~\ref{fig:dsigdq2_renscaleall}$a$ is the pure LO
calculation (dashed line) and the calculation complemented with parton
showers (dotted line). When parton showers are included in the
calculation, the LO cross-section increases by about $15\%$ at low
$Q^2$ and by about $10\%$ at large $Q^2$.  At low $Q^2$, the NLO
correction increases \sigmabjm~by almost a factor of $2$, toward
larger $Q^2$ the NLO correction increases \sigmabjm~by $1.6$. When
parton showers are included the NLO correction is lower,
\sigmabjm~increases by $1.75$ at low $Q^2$ and by $1.5$ at large
$Q^2$.

For \bquark production in DIS there are three possible choices for the
hard scale ${\cal Q}^2$: $m_b^2$, $Q^2$ and the transverse momentum of
the \bquark or of the associated jet in the Breit frame.  The
transverse momentum can either be defined to be the transverse
momentum of the \bquark $p_{T,b}$ or the maximal transverse momentum
of the \bquark or the $\bar{b}$-quark $p_{T,b, {\rm max}}$.  In
principle, any function of these scales is a possible choice.  If one
of the possible scales is much larger or smaller than the others, but
still sizable, it can be expected that large logarithms of the form
$\log{({\cal Q}_1^2/{\cal Q}_2^2})$ appear in the calculation. These
logarithms have to be resummed to make the calculation reliable.

The relative difference of the cross-section calculated with different
scale choices, i.e.\ $(\sigmabjm({\cal Q}_1^2) - \sigmabjm({\cal
  Q}_0^2))/\sigmabjm({\cal Q}_0^2),$ is shown in
figure~\ref{fig:dsigdq2_renscaleall}$b$ as a function of $Q^2$.  The
default scale choice is ${\cal Q}_0^2 = p_{T,b}^2 + 4 m_b^2$. Shown as
examples are ${\cal Q}_1^2 = Q^2 + 4 m_b^2$ (solid line), ${\cal
  Q}_1^2 = 4 m_b^2$ (dashed line), ${\cal Q}_1^2 = p_{T,b}^2 + Q^2 + 4
m_b^2$ (dotted line), ${\cal Q}_1^2 = p_{T,b, {\rm max}}^2 + 4 m_b^2$
(dashed-dotted line).  All studied possible scale choices lead to
cross-section predictions within the uncertainties of the default
choice.  The choices ${\cal Q}_1^2 = p_{T,b}^2 + Q^2 + 4 m_b^2$ and
${\cal Q}_1^2 = p_{T,b, {\rm max}}^2 + 4 m_b^2$ give exactly the same
result.  The cross-section calculated for ${\cal Q}_1^2 = 4 m_b^2$ is
$10 \%$ higher over the full $Q^2$ region, the one calculated for
${\cal Q}_1^2 = Q^2 + 4 m_b^2$ is $10 \%$ higher at lower $Q^2$, but
only $2\%$ higher at large $Q^2$.

\looseness=-1 The NLO QCD cross-section prediction is therefore more or less
independent on the choice of the hard scale. One way to decide which
scale choice should be used as the default, is to study the
uncertainties introduced by the various possible scale choice.
Figure~\ref{fig:dsigdq2_renscale1} shows the cross-section uncertainty
for the six studied scale choices as a function of $Q^2$. For all
choices the total uncertainties stay within $\pm 20\%$. The
uncertainty is larger at low $Q^2$ for all choices and decreases
toward larger $Q^2$. When the transverse \bquark momentum is included
in the definition of ${\cal Q}^2$, the uncertainty is in general
smaller. For instance, for ${\cal Q}_1^2 = Q^2 + 4 m_b^2$ the
uncertainty is around $\pm (10-20)\%$, while for ${\cal Q}_1^2 =
p_{T,b}^2 + Q^2 + 4 m_b^2$ the uncertainty is around $\pm
(10-15)\%$. For ${\cal Q}_1^2 = 4 m_b^2$ the uncertainty is $\pm
(15-20)\%$, while for ${\cal Q}_1^2 = p_{T,b}^2 + 4 m_b^2$ the
uncertainty is only $\pm (10-15) \%$. No big difference is seen
between choices involving $p_{T,b}$ or $p_{T,b, {\rm max}}$.  The
smallest scale uncertainty is obtained for ${\cal Q}_1^2 = p_{T,b}^2 +
4 m_b^2$ or ${\cal Q}_1^2 = p_{T,b, {\rm max}}^2 + 4 m_b^2$.  The
choice ${\cal Q}_1^2 = p_{T,b}^2 + 4 m_b^2$ is therefore proposed as
default.

{\renewcommand\belowcaptionskip{1em}
\FIGURE[t]{\epsfig{file=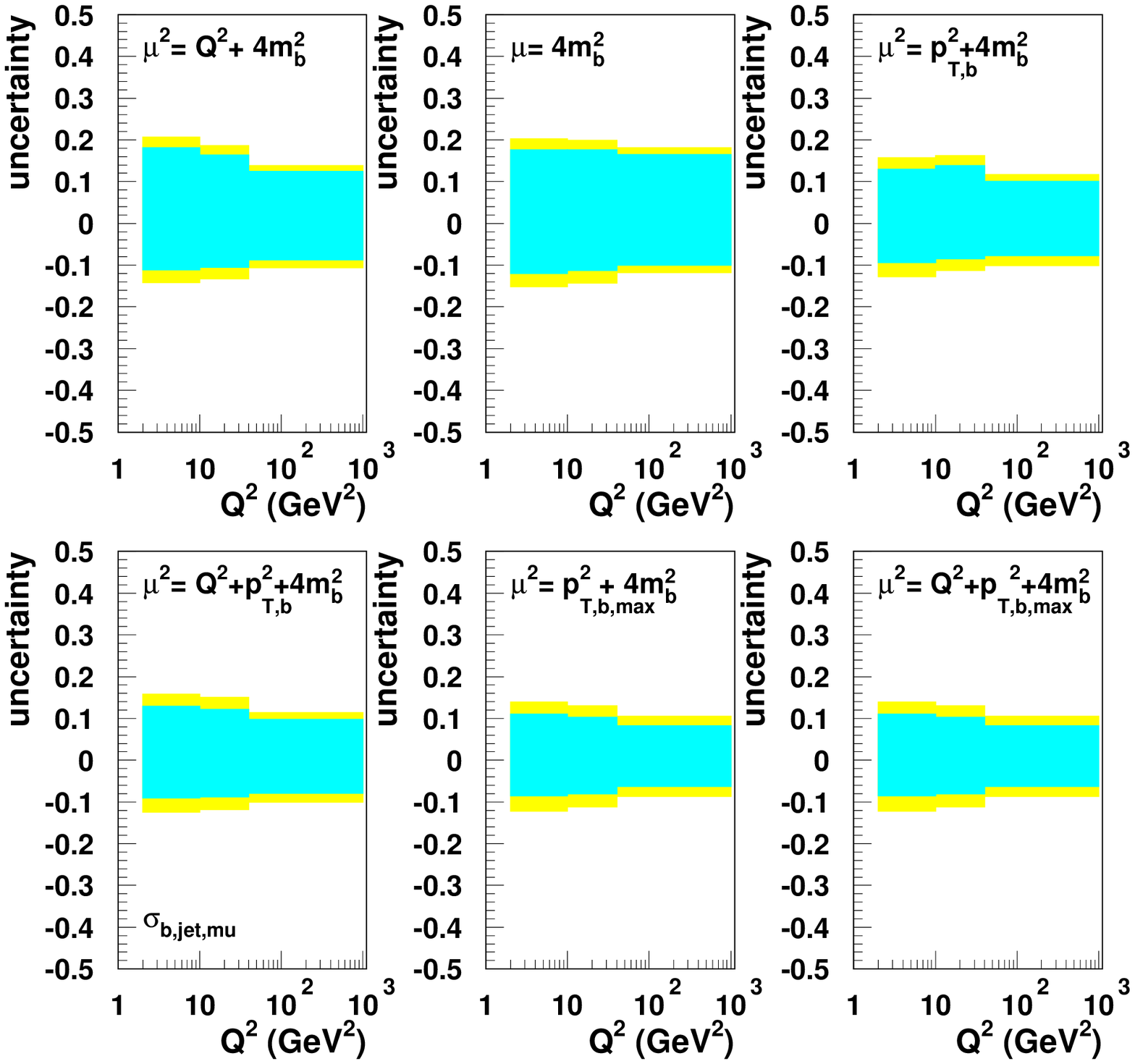,width=.8\textwidth}%
\begin{picture}(0,0) 
\put(-310,165){($a$)}
\put(-200,165){($b$)} 
\put(-90,165){($c$)}
\put(-310,0) {($d$)}
\put(-200,0) {($e$)}
\put(-90,0) {($f$)}  
\end{picture} 
\caption{Uncertainty of the \bquark cross-section \sigmabjm~ as a
  function of $Q^2$ for different choices of the renormalisation and
  factorisation scales.  The inner band represents the renormalisation
  and factorisation scale uncertainties and the outer band in addition
  the uncertainty due to the \bquark mass.  The scales are explained
  in the text.\label{fig:dsigdq2_renscale1}}}}

\FIGURE[t]{\epsfig{file=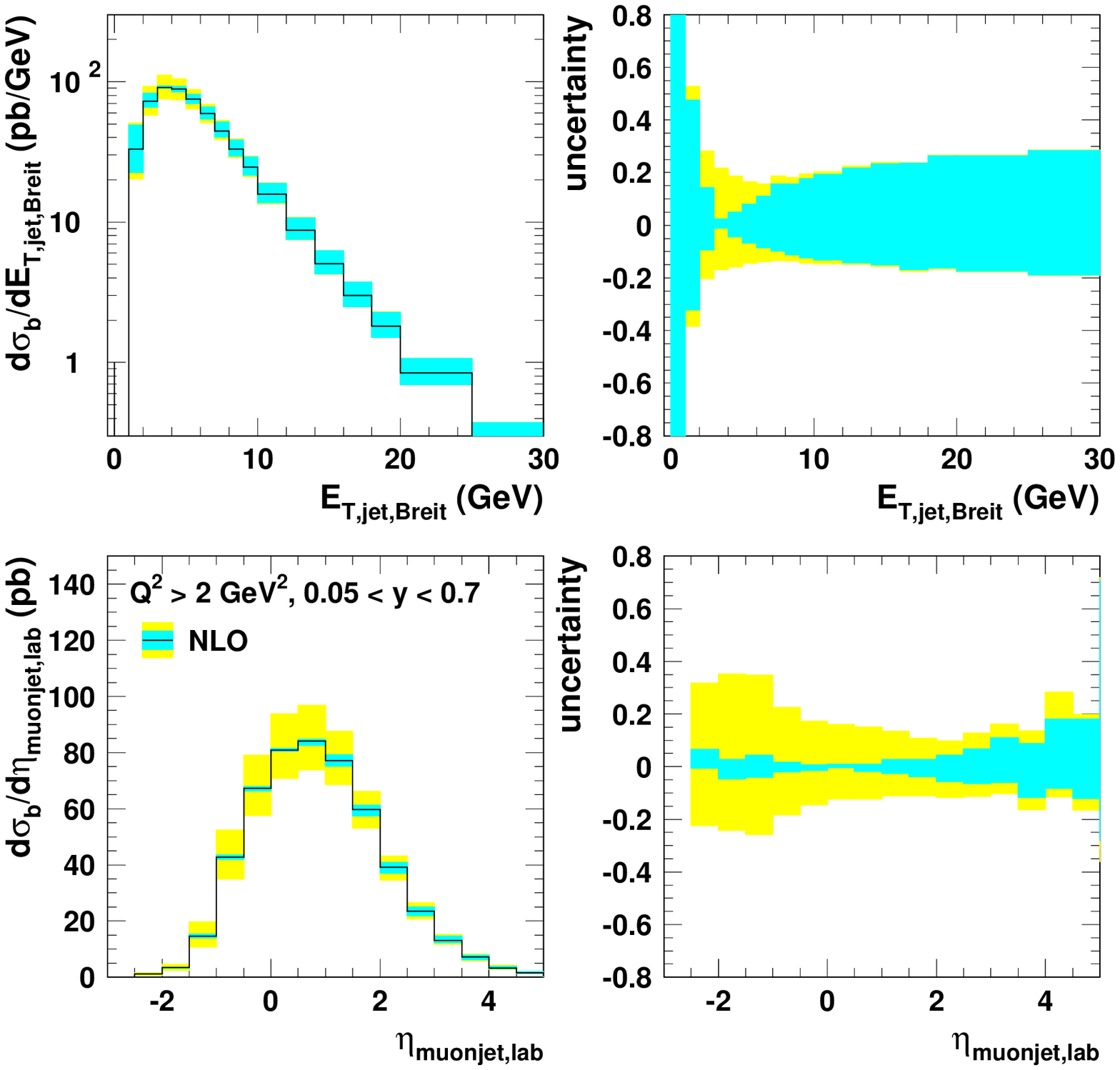,width=.8\textwidth}%
\begin{picture}(0,0) 
\put(-320,170){($a$)}
\put(-150,170){($b$)} 
\put(-320,0) {($c$)}
\put(-150,0) {($d$)} 
\end{picture} 
\caption{The cross-section \sigmab\ as a function of the jet with the
  highest transverse energy \etjet ($a$) and the pseudo-rapidity
  \etamujet\ of the jet associated with the muon ($c$).  The inner band
  represents the renormalisation and factorisation scale uncertainties
  and the outer band in addition the uncertainty due to the \bquark
  mass.  The uncertainty, i.e.\ the ratio of the cross-section
  calculated with the default settings over the one with varied
  parameters is shown in ($b$) and ($d$).  The renormalisation and
  factorisation scale have been set to $\mu^2 = p_{t,b}^2+4
  m_b^2$.\label{fig:dsigdetab_ptb24m2}}}

Figure~\ref{fig:dsigdetab_ptb24m2} shows the inclusive \bquark
cross-section \sigmab~(left) as a function of the transverse energy of
the jet with the highest transverse energy ($E_{T,jet,Breit}$) in the
event and as a function of the pseudo-rapidity of the jet which
contains the muon from the semi-leptonic \bquark decay. The scale is
set to $\mu^2 = p_{T,b}^2 + 4 m_b^2$.  In the right part of
figure~\ref{fig:dsigdetab_ptb24m2} the uncertainty of the NLO
calculation is explicitly shown. It is interesting that the scale
dependence of the cross-section varies as a function of the shown
observables. It is most pronounced toward larger transverse jet
energies. This is the case where the transverse jet energy and
consequently the transverse \bquark momentum are very different from
the \bquark mass. A large scale dependence is also seen in the forward
direction.  For low $E_{T,jet,Breit}$, where $E_{T,jet,Breit} \approx
m_b$, the scale dependence is rather small (down to $\pm 5\%$).  This
is the region where most of the inclusive \bquark events are produced.
Therefore the small scale dependence of the inclusive cross-section
\sigmab~which was discussed in the context of
figure~\ref{fig:dsigvsmu}, seems to be purely accidental.

The total uncertainty including the \bquark mass is about constant
namely around $\pm 20 \%$. Only in the large $E_{T,jet,Breit}$ region
the uncertainty is increased up to about $30 \%$.  In the region where
the scale total uncertainty is small, the uncertainty due to the
\bquark mass is large.

\looseness=-1 The increase of the uncertainties toward low
$E_{T,jet,Breit}$ is most probably due to an infra-red sensitive phase
space region~\cite{sens1,sens2}.  In fact, the \bquark cross-section
steeply falls toward $E_{T,jet,Breit} \to 0$ and even gets slightly
negative in the extreme limit (not shown). This reflects an incomplete
cancellation of the (positive) divergences of the real infra-red
parton emissions and the (negative) divergences of the virtual
contributions.  The only possibility that no hard jet is produced in
the event, is the configuration where the \bquark and the
$\bar{b}$-quark are scattered exactly in (or anti-parallel to) the
direction of the proton.

Note, that the $20\%$ total uncertainty on the inclusive cross-section
\sigmab~is larger than the one on the \sigmabjm~cross-section which is
overall only about $15\%$ for $\mu^2= p_{t,b}^2+4 m_b^2$ (see
figure~\ref{fig:dsigdq2_renscale1}$c$).  The differential cross-section
for the case \sigmabjm~is shown in figure~\ref{fig:dsigdetab_bjm}.
The uncertainties exhibit a similar behaviour than in the inclusive
case, i.e.\ the scale dependence gets larger toward larger transverse
jet energies and toward the forward region, while the \bquark mass
uncertainty is largest at low transverse jet energies and in the
backward region.  However, the overall uncertainty is lower than in
the inclusive case.  At low $E_{T,jet,Breit}$, it is only about $\pm
(10 - 15) \%$ and increases to about $ \pm 20 \%$ toward larger
$E_{T,jet,Breit}$. The uncertainties in the backward and forward
region toward the end of the detector acceptance are about the same in
the case of \sigmab~ and \sigmabjm.

\FIGURE[t]{\epsfig{file=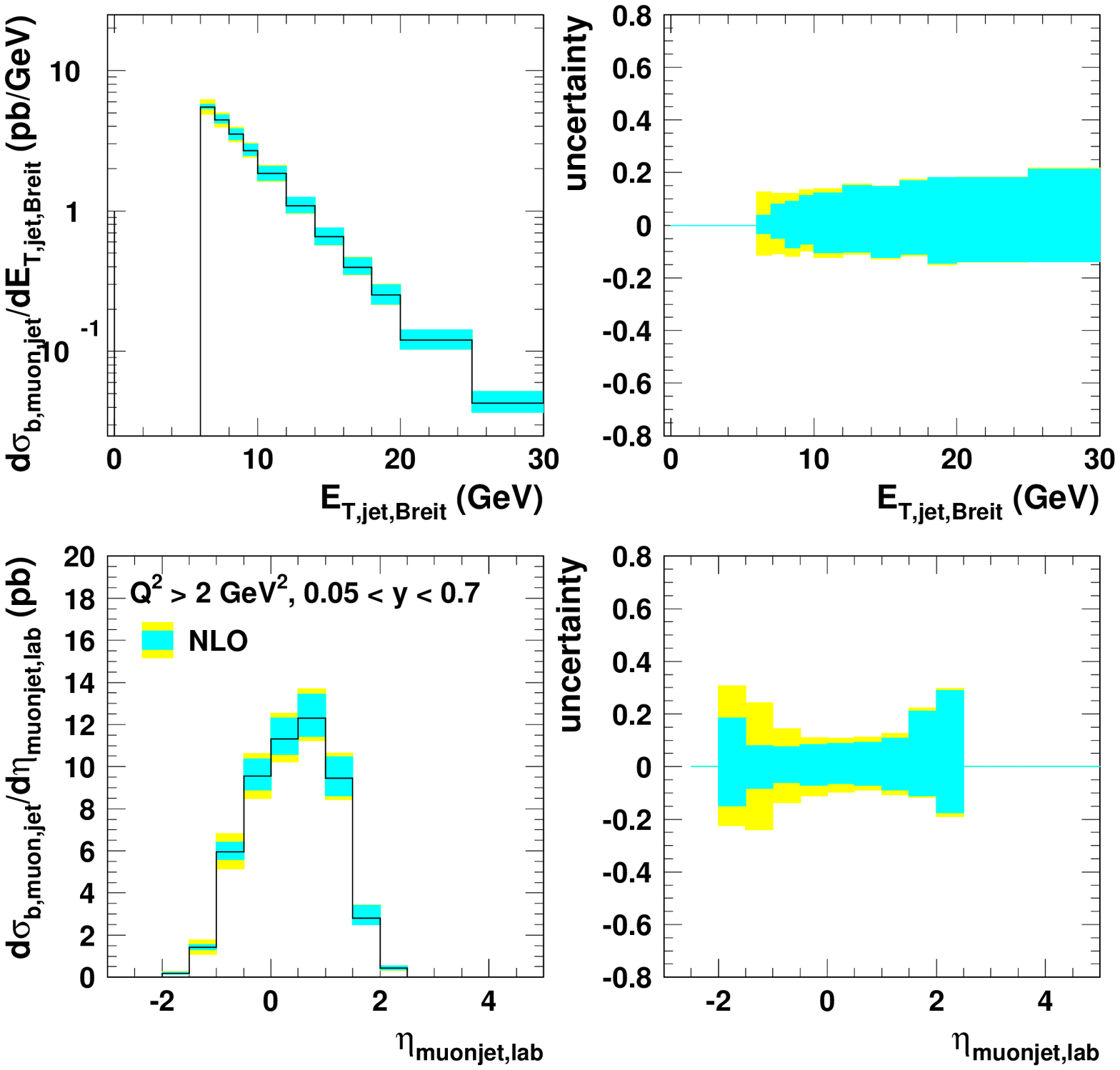,width=.8\textwidth}%
\begin{picture}(0,0) 
\put(-320,170){($a$)}
\put(-150,170){($b$)} 
\put(-320,0) {($c$)}
\put(-150,0) {($d$)} 
\end{picture} 
\caption{Same as figure~\ref{fig:dsigdetab_ptb24m2}, but for the
  measurable cross-section \sigmabjm.\label{fig:dsigdetab_bjm}}}

\subsection{Parton density uncertainties}

The calculation of cross-sections for observable processes generally
involve the convolution of the perturbatively calculable coefficient
functions corresponding to the hard subprocess with the
non-perturbative parton densities which describe the structure of the
proton. The parton densities absorb the collinear divergences of
initial state parton radiation appearing in the perturbative
calculation. The boundary defining which contributions are
perturbatively calculated and which are absorbed in the parton
densities is given by the factorisation scale.  The parton densities
can be extracted in a fit procedure by comparing cross-sections
calculated by convolution them with the coefficient functions to
data. Usually, a variety of data such as inclusive DIS, Drell-Yan,
prompt photon, $W$-boson and inclusive jet production in $ep$ or $p
\bar{p}$ collisions etc. are used to optimally constrain the parton
densities.  Recently, significant progress has been made in
determining in addition also the uncertainties on the parton
densities. This is achieved by a careful error propagation in the fit
procedure.

To estimate the impact of the limited knowledge of the proton parton
densities on the calculation of the \bquark cross-section, we use the
recent NLO QCD fit performed by the ZEUS
collaboration~\cite{Chekanov:2002pv}.  Since the large part of \bquark
event are initiated by gluons, the cross-section uncertainty is mainly
due to the gluon density for gluon energy fractions $x_g$ in the
range\footnote{About $90\%$ of all \bquark events have energy
  fractions $x_g$ in this range. Due to kinematic reasons, values $x_g
  < 10^{-3}$ can not be reached. About $10\%$ of the events have $x_g
  > 10^{-1}$. In some case also very large $x_g$ values are reached.}
$10^{-3} - 10^{-1}$.  The hard scale where the parton densities are
probed lies typically within $10 - 100\GeVsq$.

\FIGURE[t]{\epsfig{file=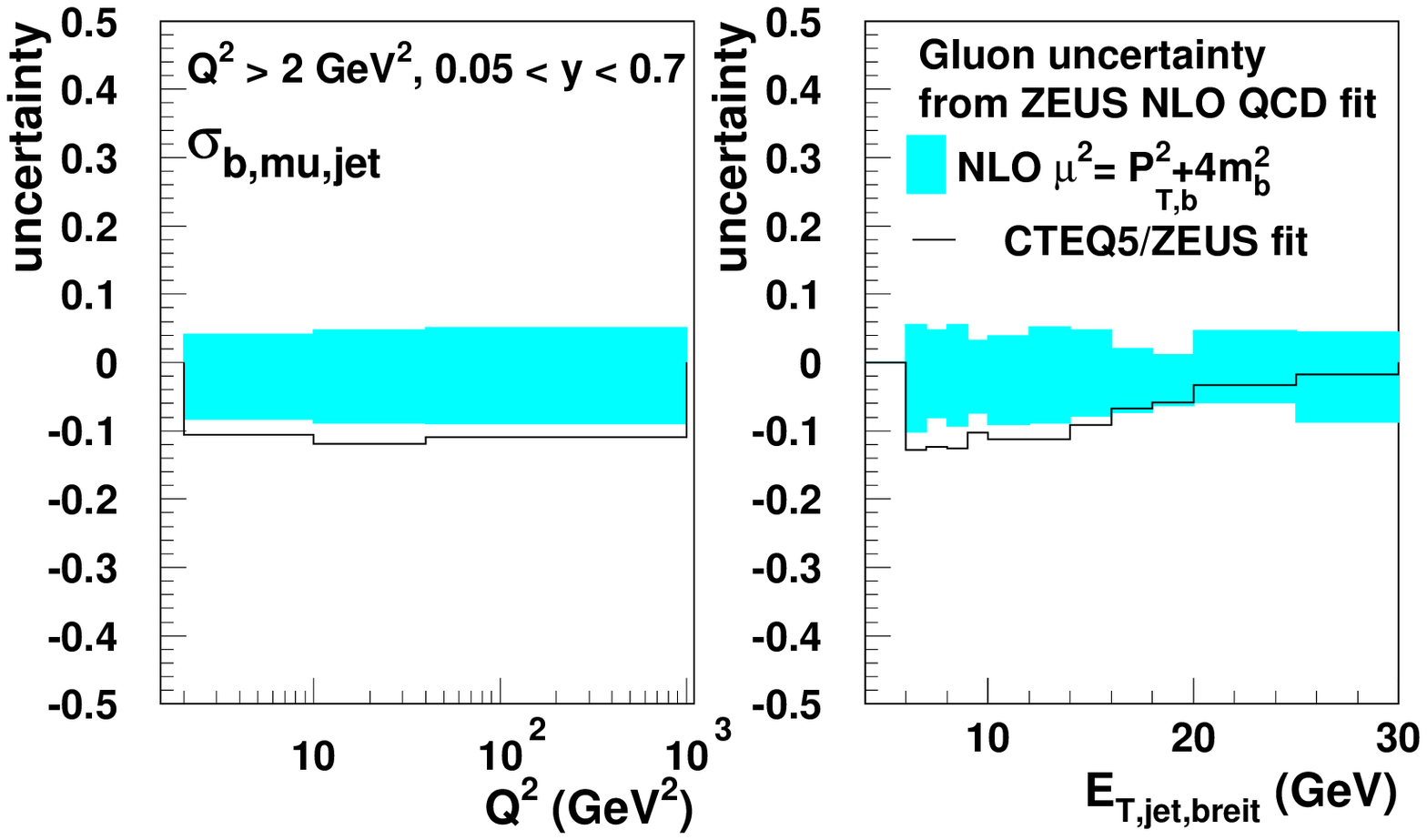,width=.8\textwidth}%
\begin{picture}(0,0) 
\put(-320,0){($a$)}
\put(-150,0){($b$)} 
\end{picture} 
\caption{Uncertainty on the \sigmabjm\ cross-section introduced by the
  uncertainty on the gluon density as a function of $Q^2$ ($a$) and
  \etjet.  The inner band represents the renormalisation and
  factorisation scale uncertainties and the outer band in addition the
  uncertainty due to the \bquark mass.  Shown as line is the ratio of
  the central ZEUS NLO QCD fit and the CTEQ5F4
  parameterisation.\label{fig:gluon_uncertainty}}}

The uncertainty of the measurable cross-section \sigmabjm~is shown as
a function of $Q^2$ in figure~\ref{fig:gluon_uncertainty}$a$ and as a
function of $E_{T,jet,Breit}$ in
figure~\ref{fig:gluon_uncertainty}$b$. The band indicates the
cross-section uncertainty due to the parton densities. The solid line
gives the ratio of the cross-section calculated with the central ZEUS
parton densities and the CTEQF4 parton densities. The cross-section
calculated with CTEQ parameterisation is overall about $10\%$
lower. The smaller cross-section is only observed at low
$E_{T,jet,Breit}$, where the bulk of the events are, for high
$E_{T,jet,Breit}$ the two different parton density parameterisations
give about the same results. The cross-section difference is not
bigger than the uncertainty determined for the ZEUS parameterisation
which is not bigger than $\pm (5-10)\%$.

\subsection{Fragmentation function uncertainties}

The uncertainty due to the fragmentation function is estimated by
changing possible choices of the fragmentation functions and their
free parameters. Since care has to be taken that the fragmentation
function is well matched to the perturbative calculation (see
section~\ref{sec:b_frag}), the estimated uncertainty is probably
bigger than the real one.

\FIGURE[b]{\epsfig{file=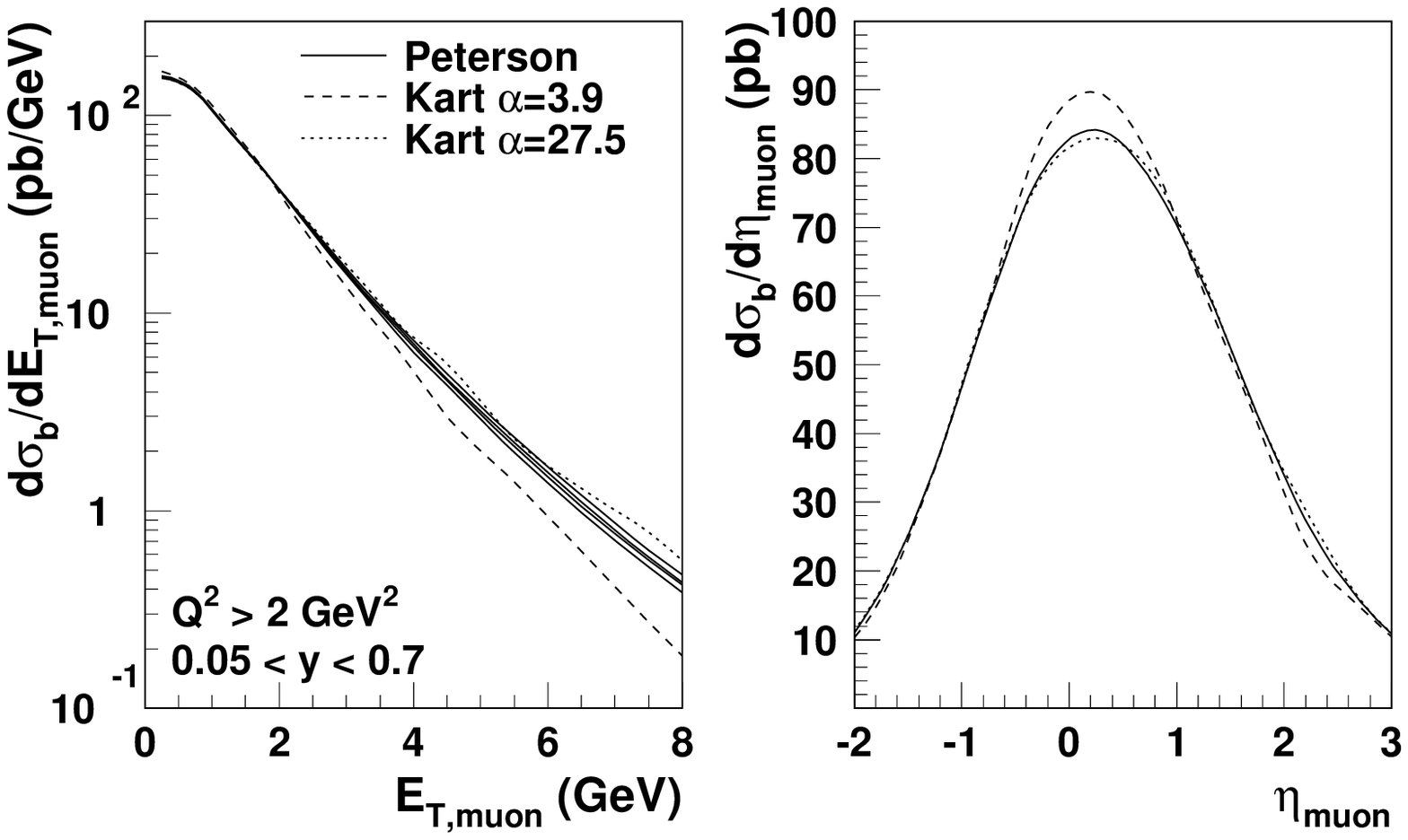,width=.75\textwidth}%
\begin{picture}(0,0)
\put(-300,0){($a$)}
\put(-130,0){($b$)} 
\end{picture} 
\caption{The \bquark cross-section \sigmab~as a function of the muon
  transverse energy ($a$) and pseudo-rapidity ($b$) calculated with
  different fragmentation functions. Shown as band is \sigmab~
  calculated with the Peterson fragmentation
  function~\cite{pr:d27:105} for $\epsilon=0.001-0.005$ and the
  parameterisations of Kartvelishvili et
  al.~\cite{Kartvelishvili:1978pi} using the SLD fit parameter and the
  one obtained by a fit to the fourth moment of the \bquark
  fragmentation function measured in $e^+ e^-$
  collisions~\cite{Cacciari:2002pa,Cacciari:2002gn}.\label{fig:fragfun_uncert}}}

Figure~\ref{fig:fragfun_uncert} shows the transverse energy and
pseudo-rapidity distribution of the muon for the inclusive case. Shown
as band is the NLO QCD cross-section \sigmab~calculated with the
Peterson fragmentation function~\cite{pr:d27:105} with different
parameters, i.e.\ $0.001 \le \epsilon \le 0.005$.  The dashed line
indicates the cross-section obtained with the parameterisation of
Kartvelishvili et al.~\cite{Kartvelishvili:1978pi} using the SLD fit
parameter~\cite{Abe:2002iq} and the dotted line is obtained by using
the fit parameter parameter proposed in
refs.~\cite{Cacciari:2002pa,Cacciari:2002gn}.  While for low transverse
muon energies all cross-section agree, noticeable difference are found
for $E_{T,muon} \gsim 2 \GeV$.  No big differences are found in the
pseudo-rapidity distribution of the muon.

Since most of the \bquark events produce muons with low transverse
energies, the total measurable cross-section \sigmabjm~varies only by
about $10\%$ when different $\epsilon$ parameters are used for the
Peterson fragmentation functions and by $5\%$ when different
fragmentation functions are used.

When the \bquark cross-section \sigmab~is calculated in leading order,
the $E_{T,muon}$ spectrum agrees with the one calculated with RAPGAP
when $\epsilon = 0.009$ is used.  The pseudo-rapidity distribution,
however, always peaks more forward.

\section{QCD model uncertainties}
\label{sec:model}
\subsection{Extrapolations uncertainties due to QCD models}

QCD models calculating the leading order \bquark cross-section using
the leading order matrix elements and models to implement higher order
radiation are the only calculations presently available that provide
the full hadronic final state of an event.  Therefore they are the
only available tools to study the detector response and to determine
the detector acceptance.  Often they are also used to extrapolate from
the measured cross-section to a more inclusive cross-section.

\TABLE[b]{\begin{tabular}{|c|cccc|}
\hline
Model  & \sigmab ~({\rm pb}) & \sigmabj ~({\rm pb}) & \sigmabm ~({\rm pb}) & 
\sigmabjm ~({\rm pb}) \\ 
\hline
NLO     & $597.6$  & $230.4$ & $52.7$ & $ 26.1$ \\
LO      & $449.4$  & $113.6$ & $40.8$ & $ 14.4$ \\ 
ME+PS   & $475.2$  & $164.7$ & $38.7$ & $ 17.1$ \\
ME only & $475.2$  & $123.3$ & $38.5$ & $ 13.8$ \\
CDM     & $475.2$  & $163.5$ & $37.0$ & $ 16.3$ \\
ME$^*$+CCFM & $822.0$  & $364.4$ & $65.8$ & $ 35.5$ \\\hline
\end{tabular}
\caption{Calculated \bquark cross-sections for the four studied definitions. 
The calculations are made with $\mu^2 = p_{T,b}^2 + 4  m_b^2$.
See the text for more explanations.\label{tab:cross-section_models}}}

In table~\ref{tab:cross-section_models} the cross-section obtained
using the QCD models is compared to the LO and NLO calculation.  {\rm
  ME+PS} denotes the parton shower model, CDM the colour dipole model
and {\rm ME$^*$+CCFM} combines the off-shell matrix elements with
initial parton radiation based on the CCFM evolution equations. The
physics content of these QCD models is explained in more detail in
section~\ref{sec:mcgen}.

For the inclusive \bquark cross-section \sigmab~the same result is
obtained for the {\rm ME+PS} and for the CDM model. Since the
inclusive cross-section does not depend on the treatment of the
hadronic final state this is also expected. It it, however,
interesting that the \bquark cross-section requiring a jet or a muon
in the final state, are also very similar for both models.  The
inclusive cross-section \sigmab~is $25\%$ lower than the NLO QCD
result. This difference between the QCD model and the NLO QCD
calculation increases to about $40\%$ when a jet or a muon is required
in the cross-section definition. While for \sigmab~the QCD models
agree with the LO QCD calculation, they lead to a $20\% - 40\%$ bigger
cross-section, when a jet is required. The increase is due to the
inclusion of the parton showers.

Since in the {\rm ME$^*$+CCFM} model a different gluon density in
addition to the different parton radiation pattern and the off-shell
matrix elements is used, the inclusive \bquark calculated
cross-section can be different.  The {\rm ME$^*$+CCFM} model gives a
cross-section which is $2-3$ times higher than the one calculated with
the {\rm ME+PS} or CDM models and about $20\% - 40\%$ higher than the
NLO QCD result.

When the QCD models are used to remove the jet in the cross-section
definition, i.e quoting \sigmabm~instead of \sigmabjm, the following
extrapolation factor has to be calculated:
\begin{eqnarray}
\sigma_{\rm b}^{\rm data} = \sigma_{\rm b,jet,muon}^{\rm data} 
\cdot
\frac{\sigma_{\rm b}^{\rm MC}}{\sigma_{\rm b,jet,muon}^{\rm MC}}\,. 
\nonumber
\end{eqnarray}

For MEPS and CDM this extrapolation factor is about $2.3$. However,
for NLO it is only $2.0$ and for {\rm ME$^*$+CCFM} it is $1.8$. Since
it is not known, which of the calculation gives the correct result,
the jet extrapolation introduces a model uncertainty of about $20\%$.
If one, for instance, assumed that the experimental measurement would
give \sigmabjm$ = 35.5 \pb$ as predicted by CASCADE and one 
used the MEPS Monte Carlo simulation to quote a cross-section
\sigmabm, the result would be $\sigmabm=78.1\pb$. This is, however,
about $20 \%$ higher than the correct result.  Moreover, the
extrapolation factor is dependent on the choice of the renormalisation
and factorisation scales. This is illustrated for the NLO QCD
calculation in figure~\ref{fig:dsigvsmu}.  The ratio of the
dashed-dotted and the dotted line is the extrapolation factor needed
to correct for the jet acceptance. The extrapolation factor varies
from $1.8$ obtained for scale choice of $\xi=0.1$ to $2.25$ for
$\xi=10$, which corresponds to a change by $20\%$.

Sometimes the jet definition is kept in the cross-section definition,
but the requirement of a muon is removed in order to get a
cross-section definition where no fragmentation model is involved in
the NLO QCD cross-section calculation. The idea of such a procedure is
that the complex fragmentation model as implemented in the QCD models
are more reliable than the very simple one used in the NLO QCD
calculation.\footnote{Even if this idea sounds reasonable one has to
  be careful, since the different partonic final state can lead to
  different distributions of the variables defining the muon
  kinematics.}  If the muon requirement is removed from the
cross-section definition, the following extrapolation factor has to be
calculated:
\begin{eqnarray}
\sigma_{\rm b, jet}^{\rm data} = 
\sigma_{\rm b, jet, muon}^{\rm data} \cdot
\frac{\sigma_{\rm b, jet}^{\rm MC}}{\sigma_{\rm b, jet, muon}^{\rm MC}} 
\nonumber
\end{eqnarray}
The correction factor calculated with MEPS, CDM or {\rm ME$^*$+CCFM}
is about $10$. The one calculated with NLO QCD is $8.8$. The
introduced model uncertainty is about $10\%$. The correction factor is
rather large, only $10\%$ of the quoted cross-section result is
actually measured in the detector. However, part of the correction is
due to the muon branching fraction which is well known.  About $75\%$
of the muon in the rapidity range measured by the detector
$-1.7<\eta_{\rm muon}<1.3$ and about $30\%$ of the muons satisfy
$p_{\rm muon} > 2$\GeV.

\FIGURE{\epsfig{file=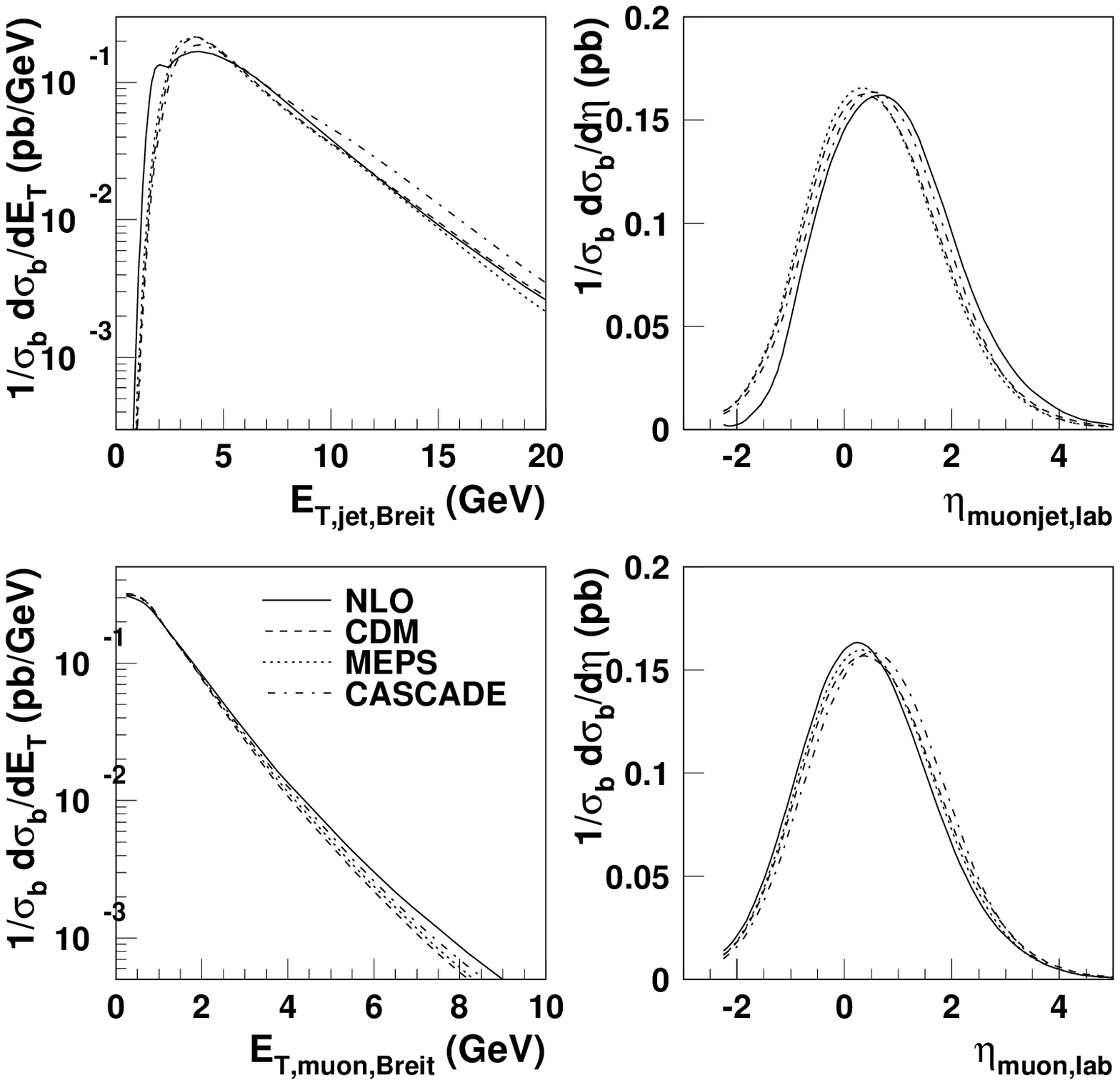,width=.8\textwidth}%
\begin{picture}(0,0) 
\put(-320,170){($a$)}
\put(-150,170){($b$)} 
\put(-320,0) {($c$)}
\put(-150,0) {($d$)} 
\end{picture} 
\caption{Comparison of the shape of the transverse energy and the
  pseudo-rapidity distribution of the jet and the muon for the inclusive
  \bquark cross-section as obtained by the NLO QCD calculation and the
  QCD Monte Carlo models.\label{fig:vglnlo}}}

The model uncertainties of the extrapolation factors are related to
the different distribution of the variables defining the
cross-section, i.e.\ the muon and jet transverse energies and
pseudo-rapidities.  The shapes of these distributions for inclusive
\bquark events are shown in
figure~\ref{fig:vglnlo}. Figure~\ref{fig:vglnlo}$a$ shows the transverse
energy of the jet with the highest $E_T$ in the Breit frame and
figure~\ref{fig:vglnlo}$b$ shows the pseudo-rapidity of the jet which is
associated with the muon.\footnote{The association of the muon and the
  jet is performed by the jet algorithm.} Figure~\ref{fig:vglnlo}$c$ and
figure~\ref{fig:vglnlo}$d$ show the transverse energy and the
pseudo-rapidity of the muon.  Given the very different treatments of
the underlying partonic final state it is remarkable that the jet and
muon distribution of the tested calculations are not too different.

The CDM and NLO QCD have a similar transverse jet energy spectrum.
The MEPS model produces less events at high $E_T$. The spectrum
calculated by {\rm ME$^*$+CCFM} is significantly harder. Less events
at low $E_T$ and more events at large $E_T$ are expected. The
pseudo-rapidity spectrum of the jet associated to the muon is similar
for the MEPS and the CDM model. NLO QCD has a spectrum which is
clearly shifted toward the forward direction.  In between these two
models lies the {\rm ME$^*$+CCFM} prediction.  In contrast to the
transverse jet energy NLO QCD exhibits the hardest transverse muon
energy spectrum.  MEPS and CDM have similar spectra.  {\rm
  ME$^*$+CCFM} lies in between. At low transverse muon energies all
calculations more or less agree. The muon pseudo-rapidity spectrum is
most shifted toward the forward region for the {\rm ME$^*$+CCFM}
prediction.  NLO QCD is most backward and CDM agrees more or less with
MEPS.

The detailed understanding of these distributions is rather
difficult. While the jet distributions are a reflection of the
different partonic final state, the muon distributions also include
hadronisation effects. It is e.g.\ remarkable that the muon
pseudo-rapidity distribution is most shifted toward the backward
region for the NLO QCD calculation, although the jet pseudo-rapidity
distribution is shifted most forward. This might be explained by the
observation made in the context of figure~\ref{fig:vgllomeps} that the
hadronisation pulls the muon forward.

\subsection{Hadronisation corrections uncertainties}

If a jet is required in the cross-section calculation, the NLO QCD
prediction should be corrected for hadronisation effects, since the
jets can only be formed by the three partons available in the final
state. Such a correction is, however, only meaningful, if the shape of
the transverse energy and pseudo-rapidity distributions of the parton
jets calculated with the QCD models is not too different from the one
of the NLO QCD calculation.

\FIGURE[b]{\epsfig{file=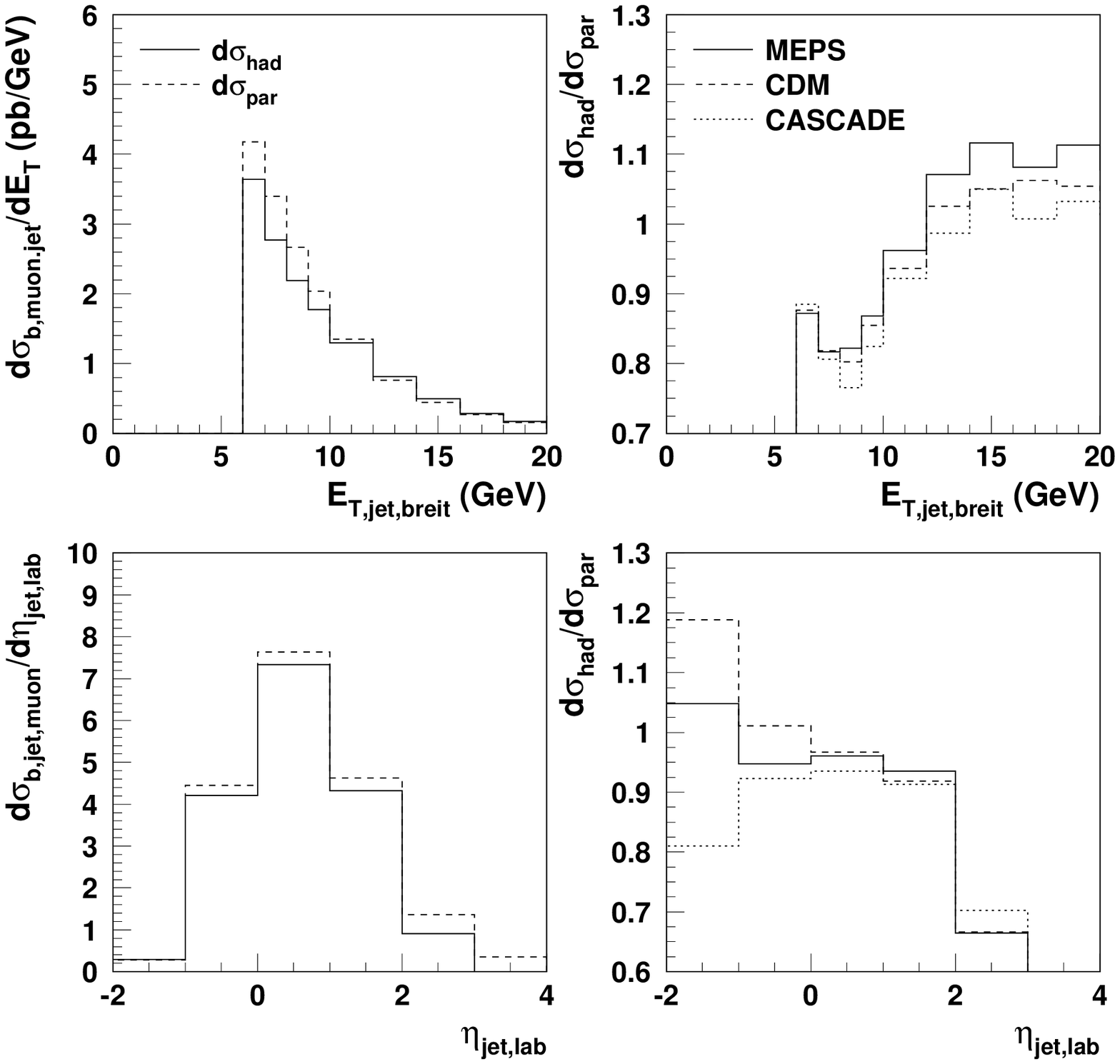,width=.8\textwidth}%
\begin{picture}(0,0)
\put(-320,170){($a$)}
\put(-150,170){($b$)} 
\put(-320,0) {($c$)}
\put(-150,0) {($d$)} 
\end{picture}
\caption{Transverse jet energy and pseudo-rapidity for the measurable
  cross-section \sigmabjm~based on the hadronic ($\sigma_{\rm had}$)
  and the partonic ($\sigma_{\rm par}$) final
  state.\label{fig:jet_hadcorr}}}

The cross-section \sigmabjm~decreases by about $10 \%$ at low $Q^2$
and by about $5\%$ at high $Q^2$, if hadronisation effects are
included.  The dependence of the hadronisation corrections on the
transverse jet energy and the jet pseudo-rapidity for the
cross-section \sigmabjm~is shown in figure~\ref{fig:jet_hadcorr}$a$ and
figure~\ref{fig:jet_hadcorr}$c$.  The solid line is the MEPS expectation
based on the hadronic final state, the dashed line is the one based on
the partonic final state.  Hadronisation effects lower the
cross-section at low $E_T$, but increase the cross-section at high
$E_T$. The hadronisation correction factor, i.e.\ $\sigma_{\rm
  had}/\sigma_{\rm par}$ is shown in figure~\ref{fig:jet_hadcorr}$b$ and
figure~\ref{fig:jet_hadcorr}$d$ for the three discussed QCD models.  The
symbol $\sigma_{\rm had}$ denotes the \bquark cross-section calculated
for hadrons, $\sigma_{\rm par}$ denotes the \bquark cross-section
calculated for partons.  The biggest model uncertainties on the
hadronisation correction is seen at large transverse jet energies and
for low jet pseudo-rapidities.  This are, however, the regions where
the cross-section is rather small.

\section{Conclusions}

The uncertainties involved in the NLO QCD calculation of \bquark
cross-sections have been estimated. Besides the inclusive \bquark
cross-section, cross-section definition requiring the muon from the
semi-leptonic \bquark decay or the jet induced by the \bquark have
been studied.  The uncertainties due to the renormalisation and
factorisation scale are about $10\%-20\%$.  The uncertainties
introduced by possible scale choices are within this margin.  In
general, the total uncertainties including the uncertainties on the
\bquark mass are about constant over the full jet transverse energy
and pseudo-rapidity range. The total uncertainties are smallest for
the cross-section where the muon and the jet are within the detector
acceptance.  When extrapolating the measurable cross-section to more
inclusive cross-section definitions large extrapolation factors can be
involved. The model uncertainties on these factors are estimated to be
$20\%$. Hadronisation corrections lead to model uncertainties of about
$10\%$.

\acknowledgments

We would like to thank M. Cacciari, H. Jung and M. Corradi for the
fruitful comments and critical reading of the manuscript.

\end{document}